\documentclass[11pt,a4paper]{article}
\usepackage[left=1in, right=1in, top=1in, bottom=1in]{geometry}
\usepackage{graphicx}
\usepackage{bm}
\usepackage{amsmath}
\usepackage{amssymb}
\usepackage{bbm}
\usepackage{mathabx}
\usepackage{caption}
\usepackage{subcaption}
\usepackage{booktabs} % for \midrule in tables
\usepackage{float}
\usepackage[final]{pdfpages}
\usepackage{listings}
\usepackage{subfiles}
\usepackage[mode=buildnew]{standalone}
\usepackage{tikz}
\usetikzlibrary{calc,arrows,patterns,intersections}
\usepackage{pgfplots}
\usepackage{xcolor}
\usepackage{hhline}
\usepgfplotslibrary{fillbetween}
\usepackage[colorinlistoftodos,prependcaption,textsize=tiny]{todonotes}
\usepackage{shellesc}
\usepackage{authblk}
\usepackage[symbol]{footmisc}
\usepackage{alphalph}

\renewcommand{\thefootnote}{\fnsymbol{footnote}}

\newcommand{\astfootnote}[1]{%
    \let\oldthefootnote=\thefootnote%
    \setcounter{footnote}{0}%
    \renewcommand{\thefootnote}{\fnsymbol{footnote}}%
    \footnote{#1}%
    \let\thefootnote=\oldthefootnote%
}

\AtBeginDocument{%
  \AtBeginEnvironment{subequations}{%
  }
}

\usepackage[backend=biber,
            style=numeric, %alphabetic
            sorting=none,
            abbreviate=false,
            dateabbrev=false,
            alldates=long]{biblatex}
\renewbibmacro*{volume+number+eid}{%
    \printfield{volume}%
    \setunit*{\addnbspace}% NEW (optional); there's also \addnbthinspace
    \printfield{number}%
    \setunit{\addcomma\space}%
    \printfield{eid}
}
\DeclareFieldFormat[article]{number}{\mkbibparens{#1}}

\renewbibmacro*{in:}{%
    {\printtext{\intitlepunct}}
}

\usepackage{color}   %May be necessary if you want to color links
\usepackage{hyperref}
\hypersetup{
    colorlinks=true, %set true if you want colored links
    linktoc=all,     %set to all if you want both sections and subsections linked
    linkcolor=blue,  %choose some color if you want links to stand out
}

\date{}

\begin{document}
\title{
    %A Pioneering Market Framework for Promoting Demand-side Flexibility in Denmark 
    %A New Ecosystem for Demand-side Flexibility Aggregators in Denmark 
    Load Shifting Versus Manual Frequency Reserve:  Which One is More Appealing \textcolor{black}{to Thermostatically Controlled Loads in Denmark}?
}
% Pioneering Market Framework for the Future Power System: the Case of Denmark
%\author{Peter Alexander Vistar Gade$^{\textrm{a}}$, Henrik Bindner, Jalal
%Kazempour}
\author[1,2]{Peter A.V. Gade\footnote{Corresponding author. Tel.: +45 24263865. \\ Email addresses: pega@dtu.dk (P.A.V. Gade), Trygve.Skjotskift@ibm.com (T. Skjøtskift), chazi@dtu.dk (C. Ziras), hwbi@dtu.dk (H.W. Bindner), jalal@dtu.dk (J. Kazempour).}}
\author[2]{Trygve Skjøtskift}
\author[1]{Charalampos Ziras}
\author[1]{Henrik W. Bindner}
\author[1]{Jalal Kazempour}
\affil[1]{Department of Wind and Energy Systems, Technical University of Denmark, Kgs. Lyngby, Denmark}
\affil[2]{IBM Client Innovation Center, Copenhagen, Denmark}
\renewcommand\Affilfont{\itshape\small}

% \date{\today}
% \maketitle
{\let\newpage\relax\maketitle}
%\tableofcontents

\section*{Abstract}

This paper investigates %how  thermostatically controlled loads can deliver flexibility 
demand-side flexibility provision in two  \textcolor{black}{distinct} forms of manual Frequency Restoration Reserve (mFRR) services and load shifting, and explores which one is financially more appealing to \textcolor{black}{Thermostatically Controlled Loads (TCLs)} in Denmark. While mFRR is an ancillary service required for the system and being bought by the system operator, load shifting is an individual act of the \textcolor{black}{TCL} in response to the variation of hourly electricity prices.  %We discuss advantages and disadvantages of the two and their appeal from a monetary point of view.
Without loss of generalization, we consider a supermarket freezer  as a representative \textcolor{black}{TCL}, and develop a grey-box model  describing its temperature dynamics using real data from a supermarket in Denmark. Taking into account price and activation uncertainties, a stochastic mixed-integer linear program is formulated to maximize the flexibility value from the freezer. %For practical reasons, we propose a linear policy to determine regulating power bids.
%, and then linearize the mFRR activation conditions. 
%For computational ease, we develop a decomposition technique, splitting the problem to a set of smaller sub-problems, one per scenario. %An out-of-sample evaluation was done on unseen Danish market price data for 2022. 
%
%We observed that load shifting was more profitable, but had a greater impact on the air and food temperatures in the freezer as opposed to mFRR that depends on the system state and bid price for activation.
%
%solution strategies for mFRR are presented: one with a simple policy and the other one with a dynamically updated policy. The proposed model accurately captures the timeline of decisions for bidding in manual frequency reserves.
%by using McCormick relaxation.
Examined on an ex-post simulation based on  Danish spot and balancing market prices in 2022, load shifting shows to be \textcolor{black}{almost as profitable as mFRR provision, although it could  be more} consequential for temperature deviations in the freezer. This indicates the need for regulatory measures by the Danish  system operator to make the \textcolor{black}{attraction of ancillary service provision  more obvious for TCLs in comparison to the load shifting alternative.} 

\section*{Keywords}

Demand-side flexibility; thermostatically controlled loads; supermarket freezer; manual frequency restoration reserves; load shifting.

\section{Introduction}

Power consumers are looking to reduce costs by any means due to sky-rocketing energy and gas prices. In Denmark, supermarkets are especially exposed due to their high energy consumption, and they are actively looking into initiatives to reduce their electricity bill. One of these initiatives is the application of demand-side flexibility, whereby \textcolor{black}{supermarket freezers as  Thermostatically Controlled Loads (TCLs)}  can shift consumption in time when spot (day-ahead) prices are comparatively lower or when the power system needs up-regulation services. There are thousands of supermarkets in Denmark willing to provide flexibility, and their freezers are especially suited for this \textcolor{black}{purpose} as they constitute a big share of their energy consumption.

IBM is currently developing a software platform, called \textit{Flex Platform}, aiming to harness flexibility from industrial and commercial consumers for bidding in ancillary service markets via a Balance Responsible Party (BRP). One large supermarket chain in Denmark has already signed up for the \textit{Flex Platform}, and another supermarket chain is currently exploring the possibility of joining.
While these supermarket chains are keen on the provision of flexibility, the estimation of their monetary benefit still remains an open question. We aim to address it by considering two potential ways for supermarkets to exploit their consumption flexibility: (\textit{i}) load shifting, and (\textit{ii}) the provision of manual Frequency Restoration Reserve (mFRR)
%\footnote{The term will be used for the remainder of the paper and is equivalent to tertiary reserves.})
services. We  develop  a stochastic optimization tool capturing the temperature dynamics and therefore consumption flexibility of supermarkets, and compute the monetary benefit of flexibility implemented via load shifting or mFRR service provision. We use two scenario generation  strategies: one using a five-day lookback on spot and balancing market prices in Denmark, and the other one using those price data in 2021. We then systematically compare these two strategies ex-post via an out-of-sample simulation based on Danish price data in 2022.

\subsection{Dilemma: Load Shifting Versus mFRR}
For power consumers who are actively looking into cost saving solutions, it is important to estimate the monetary benefit of flexibility. In Nord Pool, spot prices are announced at 2pm for the next day, so supermarkets who purchase power through a retailer at  spot prices, have a chance to shift their consumption plan for the next day, accordingly. From their perspective, it is natural to first look into load shifting as the most obvious and straightforward way to utilize flexibility. Here, load shifting simply refers to shifting consumption to another point in time, typically when spot prices are lower.

Nevertheless, load shifting, although might be attractive individually, is not necessarily in  favor of the power system.
If a significant number of consumers start to act similarly, it could be detrimental to the power system by moving the peak consumption to a different time of the day. In this case, spot prices are no longer reflecting supply-demand equilibrium, and the demand distribution over the day.
%In fact, if a significant amount of consumers move a large part of their consumption to a different time of the day, that could be detrimental to the power system by creating new and even more pronounced peaks.
Intriguingly, many industrial and residential consumers have already started shifting demand to some degree, as a response to the energy crisis.
As a result, load shifting might provide short-term monetary benefit, but it is not necessarily helping the power system in the long run.

%Instead, consumers could opt for offering ancillary services, for which they are paid for providing flexibility through an \textit{aggregator}.
As an alternative option which is certainly in favor of the power system, consumers could choose to provide flexibility by offering ancillary services through an \textit{aggregator}, for which they would be compensated. Here, we focus on frequency-supporting services, and in particular mFRR, and left a potential extension by considering more services for the future work. This service is a slow-responding reserve used to stabilize frequency in the power system after fast frequency reserves are depleted. The market for mFRR is usually operated by the national Transmission System Operator (TSO), which is Energinet in Denmark. Note that mFRR resembles load shifting in the sense that it is the largest energy reserve operated by the TSO, and can be provided by consumers via shifting their consumption in time. Likewise, generators provide mFRR by increasing their power generation.

In order for a supermarket freezer to deliver mFRR, it must be part of a larger portfolio with potentially many small assets to meet the minimum bid size requirement for entering the mFRR market. In Denmark, this minimum bid size is currently 5 MW, but it is expected to reduce to 1 MW in 2024. Furthermore, there might be additional synergy effects such as increased operational flexibility in the control response, rebound, and temperature deviations \cite{koch2011modeling}.

Without loss of generalization, we assume that one freezer delivering mFRR can be scaled to many freezers in the same way. Therefore,
%(and likewise for load shifting).
demand response for consumers participating in the mFRR market versus load shifting will  look similar. From the perspective of consumers, it is of great interest to explore the monetary benefit of mFRR versus load shifting. Similarly, it is important for TSOs to understand the incentive of consumers to provide mFRR services opposed to simply shifting load in time.

\subsection{Research questions and literature review}
Using real data from Denmark, this work aims to investigate and answer the following three research questions: (\textit{i}) how can the flexibility of a supermarket freezer as a \textcolor{black}{TCL} be characterized?
(\textit{ii}) how much monetary benefit can this \textcolor{black}{TCL} earn by the mFRR provision opposed to load shifting? and finally (\textit{iii})
what are the pros and cons of participating in the Danish mFRR market versus load shifting? Unlike the second and third questions, the first one has been addressed extensively in the literature. Therefore, the following literature review focuses on the first question, where we also differentiate our work from the current literature. To the best of our knowledge, there is no  work in the existing literature addressing the second and third questions.

%we provide a  literature review to pinpoint where new work is required, and how our work differs from the current literature. 

%Demand-side flexibility has been extensively studied.
The approach taken in each study in the literature is very much dependent on the perspective from which flexibility is utilized. Often, full knowledge and control of all assets is assumed. A prevalent implicit assumption is that the aggregator and the trading entity are the same, with an exclusive business relationship to flexible consumers \cite{gade2022ecosystem}. The incentives and investments required for delivering flexibility for the flexible consumer are often overlooked as well. Many also assume an idealized market mechanism or simply propose a new mechanism for trading flexibility.

In \cite{petersen2012eso2} and \cite{pedersen2013direct}, a complete white-box model of a supermarket refrigeration system is presented and validated against real-life data. It is also shown how such a system can provide demand response. \textcolor{black}{These two} work\textcolor{black}{s} \textcolor{black}{provide a} benchmark for any grey-box model of a supermarket refrigeration system. However, such an approach is hard to scale and requires complete knowledge of every refrigeration system. In \cite{pedersen2016improving}, a second-order model is used to model the food and air temperature in a freezer in an experimental setting. It is shown how food temperature has much slower dynamics than air temperature. In \cite{hao2014aggregate}, a simple first-order virtual battery model for a TCL is presented. This model also constitutes the starting point for modeling temperature dynamics in this work, as it has an intuitive interpretation. A similar bucket model is introduced in \cite{petersen2013taxonomy}.

In \cite{de2019leveraging}, a Mixed-Integer Linear Programming (MILP) problem  is developed to  address up- and down-regulation hours from a baseline consumption. It is assumed that the energy for down-regulation is equal to the energy not consumed when up-regulating. A similar approach is taken in our work, except we will define the use of down-regulation until the temperature state is (approximately) back to its setpoint.

\textcolor{black}{In \cite{schaperow2019simulation,chanpiwat2020using, moglen2020optimal}}, residential air condition units are modeled using up- and down-regulation blocks characterizing flexibility. The blocks are obtained from grey-box models of households \cite{siemann2013performance}. The authors show how such a block formulation can be solved using exact and stochastic dynamic programming in the context of peak shaving demand for a utility in the U.S. The demand-side flexibility then functions as a hedge towards extreme electricity prices. Although such a use case has not been relevant in Denmark yet, the estimation of flexibility blocks is a novel idea as it avoids having to explicitly integrate a physical model into an optimization problem. However, a trade-off is that the curse of dimensionality quickly makes dynamic programming computationally intractable for a large portfolio of heterogeneous demand-side assets. Furthermore, it might be difficult to assure the Markov property \cite{MarkovProperty} when reformulating the problem even slightly. Flexibility blocks are also used in \cite{bobo2018offering} for an offering strategy. Here, the flexibility blocks are derived from measurements of residential appliances in the Ecogrid 2.0 project \cite{ecogrid}.

In \cite{biegel2013information} and \cite{BiegelConstractingFlexServices}, it is described how pre-defined asset characteristics allow aggregators to utilize and contract their flexibility through an interface. This assumes that such characteristics are fully known beforehand. However, this type of information flexibility interface is still useful for aggregators when engaging with consumers with well-known assets at the point where aggregation and penetration of demand-side flexibility is a mature and prevalent business.

In \cite{biegel2013electricity}, a portfolio of residential heat pumps is modeled using a linear first-order model. Individual units are lumped together as essentially one big, aggregated heat pump as described in \cite{biegel2013lumped}. The first-order model is a grey-box model incorporating the physics, i.e., the temperature dynamics in a household. Such an approach will be used in this work as well. The authors also consider how the portfolio of heat pumps can be used for load shifting and real-time bidding in the balancing market, but do not consider the mFRR service provision. However, they assume that all the assets can be controlled continuously in a  model predictive control  setting. Temperature deviations are minimized directly in the objective function using the integrated error of total deviations. This introduces a trade-off parameter in the objective function that must be tuned to weigh the economic value versus the temperature deviation compared to the setpoint.

\textcolor{black}{Similar studies exist in the literature albeit in different settings. For example, \cite{iria2018trading} uses a two-stage stochastic optimization for prosumers to bid into tertiary reserves, i.e., mFRR. A similar approach is taken in \cite{la2021mixed} although with a focus on scalability to thousands of assets and distribution of flexibility amongst prosumers. In \cite{nitsch2021economic}, it is shown how batteries can be used to provide secondary reserves. TCLs have also been investigated for provision of primary reserves in the Nordics \cite{paridari2020flexibility}, in a microgrid \cite{mendieta2020primary}, and in the Australian power grid \cite{attarha2020network}}.

\subsection{Our contributions and outline}
To the best of our knowledge, no work has \textcolor{black}{thoroughly investigated and compared the monetary value and incentives for TCLs to} provide mFRR services versus load shifting, although both have been studied extensively. This work aims to fill that gap by taking a holistic view on the incentives from \textcolor{black}{a supermarket's} perspective in Denmark. Furthermore, when stating the objective functions for mFRR, most studies have assumed a simplified market structure, and neglect the fact that the aggregator and the consumer are not necessarily the same entity. We discuss the consequences of this assumption in relation to the market structure in Denmark \cite{gade2022ecosystem}, and in the context of mFRR and load shifting.

We provide a realistic model formulation of the mFRR bidding, taking into account the sequence of decisions needed to be made. We propose two solution strategies for mFRR bidding: (\textit{i}) A simple one learned using the Alternating Direction Method of Multipliers (ADMM) on 2021 price data which can be applied for all of 2022, and (\textit{ii}) A dynamic one which computes a new policy every day by looking at spot prices for the past five days. Real data of a supermarket freezer  in Denmark is used.

The rest of the paper is organized as follows. Section \ref{sec:monetizing_flex} describes modeling a supermarket freezer as a TCL. Thereafter, mFRR and load shifting are described in detail and their objective functions are stated. Section \ref{sec:OptimizationModel} presents the proposed stochastic optimization problem.
%Furthermore, the two solution strategies for solving the optimization problem for mFRR are introduced. 
Section \ref{sec:results} provides  results for a case study. Section \ref{sec:conclusion} concludes the paper.  \textcolor{black}{Appendix A} presents the full model formulation. Finally, \textcolor{black}{Appendix B provides a sensitivity analysis on the number of scenarios.} 

\section{Describing flexibility of a TCL}\label{sec:monetizing_flex}

There are several \textcolor{black}{distinct} ways to \textcolor{black}{exploit and} monetize flexibility from a \textcolor{black}{TCL}. In this section, we focus on mFRR and load shifting. First, we describe how to mathematically model a TCL as a flexible resource. Second, we describe how to monetize the flexibility from TCLs for mFRR and load shifting, and provide the objective functions in both cases. 
For mFRR, the objective function includes all costs and revenues for the BRP, while for load shifting, the objective function only includes the flexible demand's perspective. This approach explicitly shows the situation where flexible demand is activated without including the BRP, as this is more realistic.

\subsection{Modeling TCL as a flexible resource}
TCLs are characterized by being controlled such that the temperature is kept at a specified setpoint. Examples include heat pumps, freezers, air condition units, etc. They constitute an important part of demand-side flexibility due to  inherent thermal inertia of such temperature-driven systems \textcolor{black}{\cite{koch2011modeling,hao2014aggregate}}.

We focus on freezers only, which are a very common type of TCLs. Specifically, we focus on a single freezer display in a Danish supermarket. Freezers are characterized by a large thermal inertia due to the frozen food, which makes them suitable for flexibility \textcolor{black}{provision}. On the other hand, there is a risk of food degradation when utilizing flexibility. Therefore, it is important to model the temperature dynamics in the freezer for a realistic and risk-aware estimation of its flexibility.

The rest of the section is organized as follows. First, we visualize the measurements from a real supermarket freezer. Second, we introduce a second-order grey-box model that characterizes the supermarket freezer. Third, we validate the second-order model and show how it can be used to simulate demand response from a freezer.

\subsubsection{Supermarket freezer description}

\begin{figure}[!t]
    \centering
    \includegraphics[width=\columnwidth]{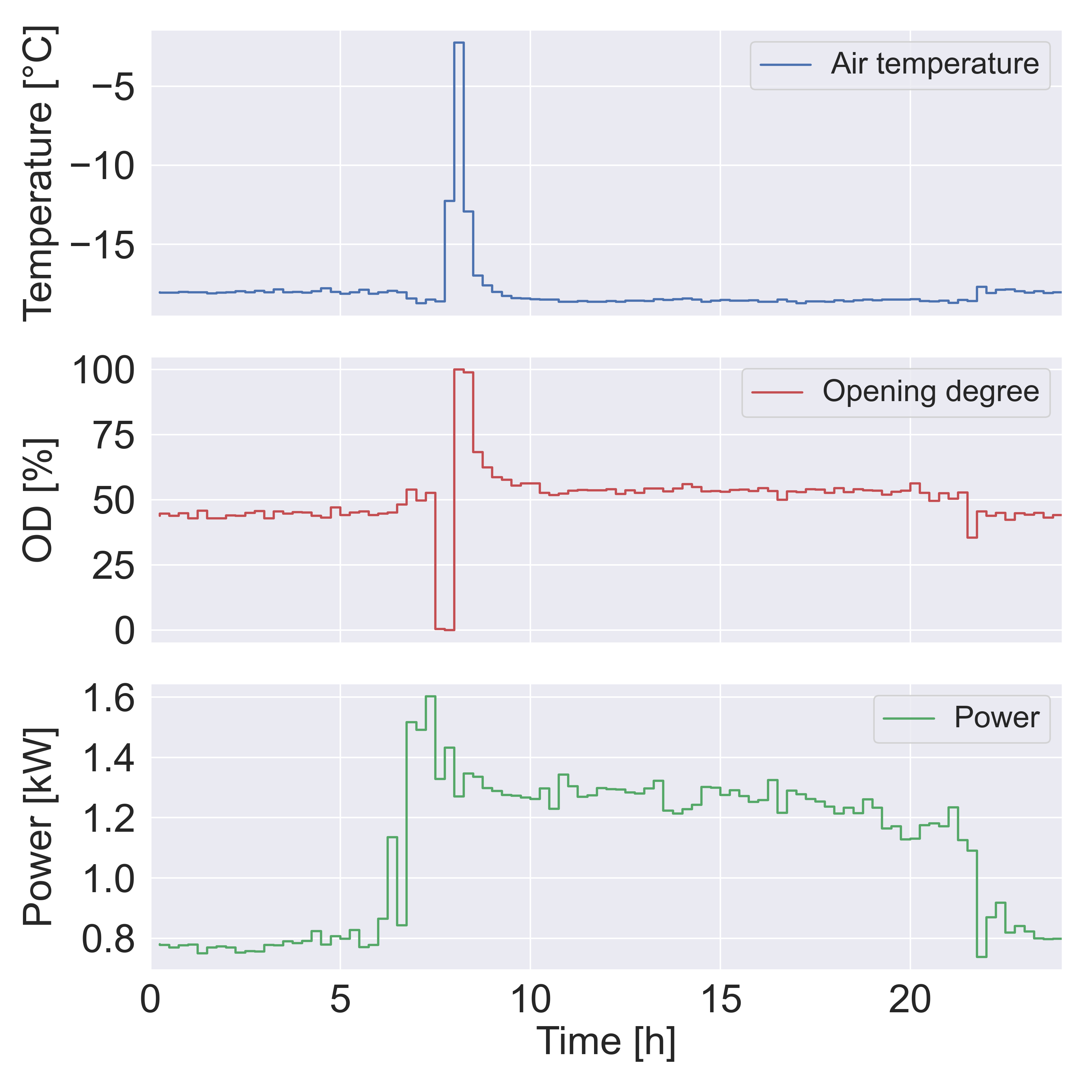}
    \caption{\textbf{Top}: temperature of a single freezer in a supermarket. \textbf{Middle}: opening degree (OD) of the freezer expansion valve. \textbf{Bottom}: electric power of the compressor rack feeding a single freezer.}
    \label{fig:chunk}
\end{figure}

We use real data from a single freezer operating in a large Danish supermarket as a case study.
In Fig. \ref{fig:chunk}, the top plot shows the 15-minute average air temperature of the freezer, whereas the middle plot depicts the opening degree of the valve.
Temperature fluctuates around its setpoint at -18 $^{\circ}$C with the exception of hours 7 and 8, where defrosting is scheduled.
While defrosting, a heating element is briefly turned on, and the expansion valve is closed, such that the flow of refrigerant stops. Afterwards, while recovering the temperature, the expansion valve is fully opened.
The electric power of the variable-speed compressor rack, scaled to one freezer is shown in the bottom plot. Since temperature dynamics are similar for all freezers, homogeneity is assumed. Hence, an equal consumption level is assumed for all freezers.

Consumption is highest during opening hours, and it is lowest during closing hours.
During opening hours, food is being replaced and customers open the display case constantly.
Furthermore, most supermarkets put additional insulation on the display cases during closing hours which reduces thermal losses.
For these reasons, there are effectively two regimes for a supermarket freezer plus a short defrosting regime.

\subsubsection{Thermal modeling of freezer}

In \cite{hao2014aggregate}, it is described how a simple TCL model can be developed. We extend it to a second-order state-space model that accounts for the thermal mass of the food, which essentially provides the flexibility in freezers:
\begin{subequations}\label{eq:2ndFreezerStateSpace}
    \begin{align}
        T^{\rm{f}}_{t+1} & = T^{\rm{f}}_{t} + dt\  \frac{1}{C^{\rm{f}}}\left(\frac{1}{R^{\rm{cf}}} (T^{\rm{c}}_{t} - T^{\rm{f}}_{t}) \right)                                                                                         \\
        T^{\rm{c}}_{t+1} & = T^{\rm{c}}_t + dt\  \frac{1}{C^{\rm{c}}}\Bigl(\frac{1}{R^{\rm{cf}}} (T^{\rm{f}}_t - T^{\rm{c}}_t) + \frac{1}{R^{\rm{ci}}} (T^{\rm{i}}_t - T^{\rm{c}}_t)                                          \notag \\ & \mspace{50mu} - \eta\  OD_t \ P_t \Bigr) + \epsilon \mathbbm{1}^{\rm{df}}_{t},
    \end{align}
\end{subequations}
where $T^{\rm{c}}_t$ is the measured air temperature in the freezer at time $t \in \{1, \ldots, J\}$, and $T^{\rm{f}}_t$ is the food temperature which is a latent unobserved state.
Parameters $C^{\rm{f}}$ and $C^{\rm{c}}$ are the thermal capacitance of the food and air in the freezer, respectively. In addition, $R^{\rm{cf}}$ and $R^{\rm{ci}}$ are the thermal resistance between food and air in the freezer, and air and indoor temperature, respectively.
There is essentially a low-pass filter of the air temperature in the freezer with the time constant $C^{\rm{f}}$ times $R^{\rm{cf}}$. Parameter $\epsilon$ represents the temperature change when defrosting, whereas $\mathbbm{1}^{\rm{df}}_{t}$ is the indicator function for when defrosting happens. Parameter $R^{\rm{ci}}$ is either one of two values, $R^{\rm{ci}, \text{day}}$ and $R^{\rm{ci}, \text{night}}$, to capture the differences between opening hours (6am to 10pm) and closing hours (10pm to 6am). The opening degree $OD_t$, indoor air temperature $T^{\rm{i}}_t$, and power $P_t$ are exogenous inputs as the opening degree is assumed to be fixed during a demand-response event, while only $P_t$ is controllable, and the indoor air temperature is unaffected by the freezers. Parameter $\eta$ is the compressor efficiency. The model is discretized with a time step of 15 minutes, therefore $dt = 0.25$ and $J = 96$.

\subsubsection{Model validation}

Using the R library of CTSM-R \cite{juhl2016ctsmr}, all parameters in (\ref{eq:2ndFreezerStateSpace}) have been estimated as given in Table \ref{tab:parameter_estimates}. Notice that the thermal capacitance of the air in the freezer is significantly smaller than the thermal capacitance of the food, indicating  that the food temperature changes comparatively slower. The thermal resistance between the food and air inside the freezer, $R^{\rm{cf}}$, is also significantly smaller than the thermal resistance between the air in the freezer and the indoor temperature in the supermarket, $R^{\rm{ci}}$, both during the day and night. This makes sense as the lid acts as a physical barrier insulating the freezer. Furthermore, the thermal resistance to the indoor air temperature is higher during the night, which means that less power is needed, as  observed in Fig. \ref{fig:chunk}.

% LONG FORMAT
% \begin{table}[!t]
%     \caption{Parameter Estimates of (\ref{eq:2ndFreezerStateSpace}).}
%     \label{tab:parameter_estimates}
%     \centering
%     \begin{tabular}[b]{|l|l|l|}
%         \hline
%         Parameter                   & Value & Unit            \\ \hhline{|=|=|=|}
%         $C^{\rm{f}}$                & 6.552 & kWh/$^{\circ}$C \\
%         $C^{\rm{c}}$                & 0.077 & kWh/$^{\circ}$C \\
%         $R^{\rm{cf}}$               & 5.010 & $^{\circ}$C/kW  \\
%         $R^{\rm{ci}, \text{day}}$   & 41.05 & $^{\circ}$C/kW  \\
%         $R^{\rm{ci}, \text{night}}$ & 61.25 & $^{\circ}$C/kW  \\
%         $\eta$                      & 1.561 &                 \\
%         $\epsilon$                  & 3.372 & $^{\circ}$C/h   \\ \hline
%     \end{tabular}
% \end{table}

% LONG FORMAT V2
\begin{table}[!t]
    \caption{Parameter Estimates of (\ref{eq:2ndFreezerStateSpace}).}
    \label{tab:parameter_estimates}
    \centering
    \begin{tabular}[b]{|l|l|l||l|l|l|}
        \hline
        Parameter                 & Value & Unit            & Parameter                   & Value & Unit           \\ \hhline{|=|=|=|=|=|=|}
        $C^{\rm{f}}$              & 6.552 & kWh/$^{\circ}$C & $R^{\rm{ci}, \text{night}}$ & 61.25 & $^{\circ}$C/kW \\
        $C^{\rm{c}}$              & 0.077 & kWh/$^{\circ}$C & $\eta$                      & 1.561 &                \\
        $R^{\rm{cf}}$             & 5.010 & $^{\circ}$C/kW  & $\epsilon$                  & 3.372 & $^{\circ}$C/h  \\
        $R^{\rm{ci}, \text{day}}$ & 41.05 & $^{\circ}$C/kW  &                             &       &                \\ \hline
    \end{tabular}
\end{table}

% WIDE FORMAT
% \begin{table}[!t]
%     \caption{Parameter Estimates of (\ref{eq:2ndFreezerStateSpace}).}
%     \label{tab:parameter_estimates}
%     \centering
%     \begin{tabular}[b]{|l|l|l||l|l|l||l||l|}
%         \hline
%         Parameter & $C^{\rm{f}}$ & $C^{\rm{c}}$ & $R^{\rm{cf}}$ & $R^{\rm{ci}, \text{day}}$ & $R^{\rm{ci}, \text{night}}$ & $\eta$ & $\epsilon$ \\ \hhline{|=|=|=||=|=|=|=|=|}
%         Value     & 6.552        & 0.077        & 5.010         & 41.05                     & 61.25                       & 1.561  & 3.372      \\ \hline
%     \end{tabular}
% \end{table}

\begin{figure}[!t]
    \centering
    \includegraphics[width=\columnwidth]{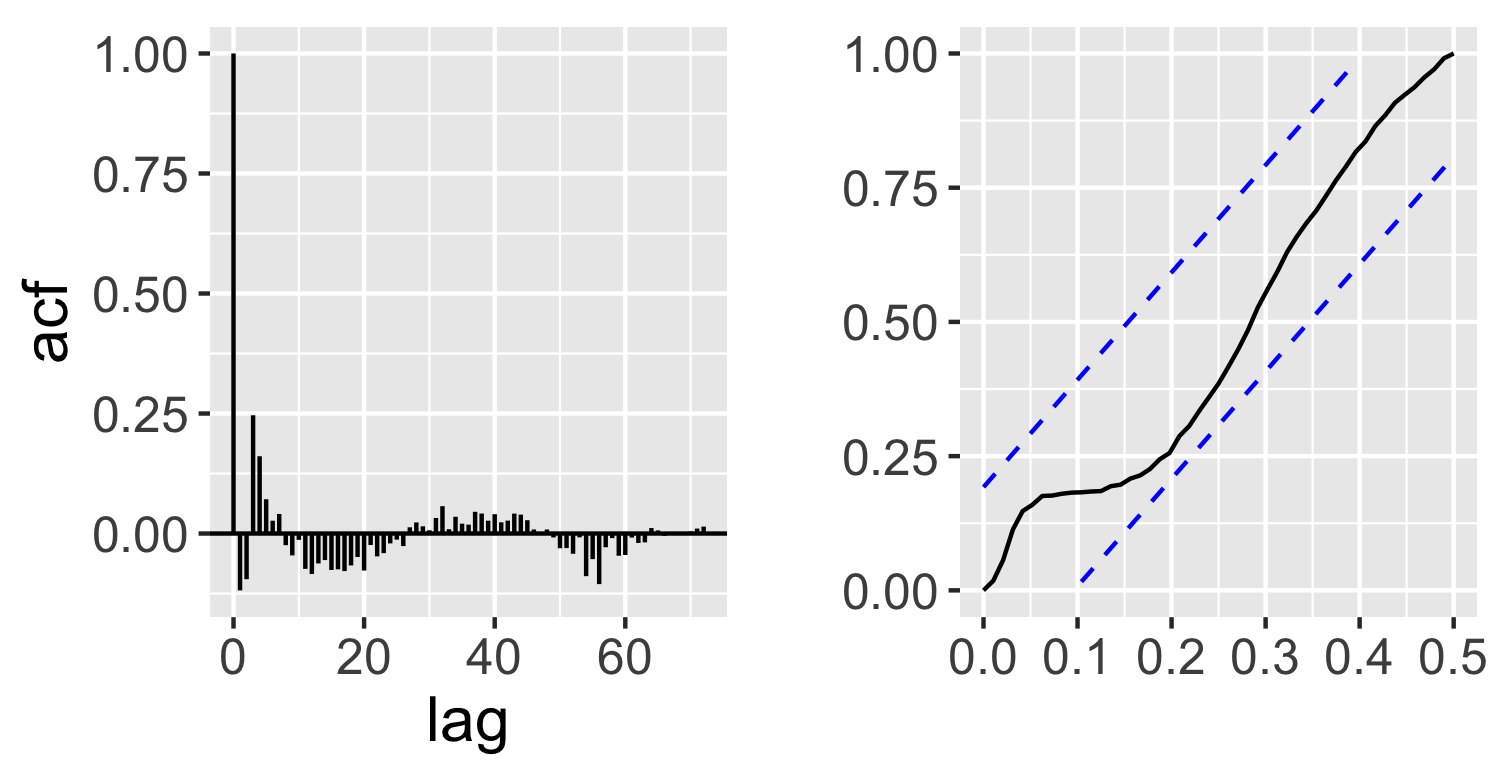}
    \caption{ Validation of the state-space model in (\ref{eq:2ndFreezerStateSpace}). \textbf{Left}: auto-correlation function (acf) of the  residuals. \textbf{Right}: cumulated periodogram of the residuals.}
    \label{fig:2ndFreezerModelValidation}
\end{figure}

The one-step residuals for the air temperature should ideally resemble white noise in order for a model to capture all dynamics observed in the data \cite{madsen2007time}. Fig. \ref{fig:2ndFreezerModelValidation} shows the auto-correlation and cumulative periodogram of the residuals. The autocorrelation shows two significant lags for lag two and seven, but looks satisfactory otherwise. Likewise,  it seems the model is able to capture most dynamics at all frequencies.
Since there are only $J = 96$ time steps, the defrosting period can result in relatively large residuals as it is difficult to capture such fast transient dynamics. Since up-regulation is not allowed during defrosting, the residuals are not too important during that period. To decrease their effect in the parameter estimation procedure, the term $ \epsilon \mathbbm{1}^{\rm{df}}_{t}$ was added to (\ref{eq:2ndFreezerStateSpace}).

\begin{figure}[!t]
    \centering
    \includegraphics[width=\columnwidth]{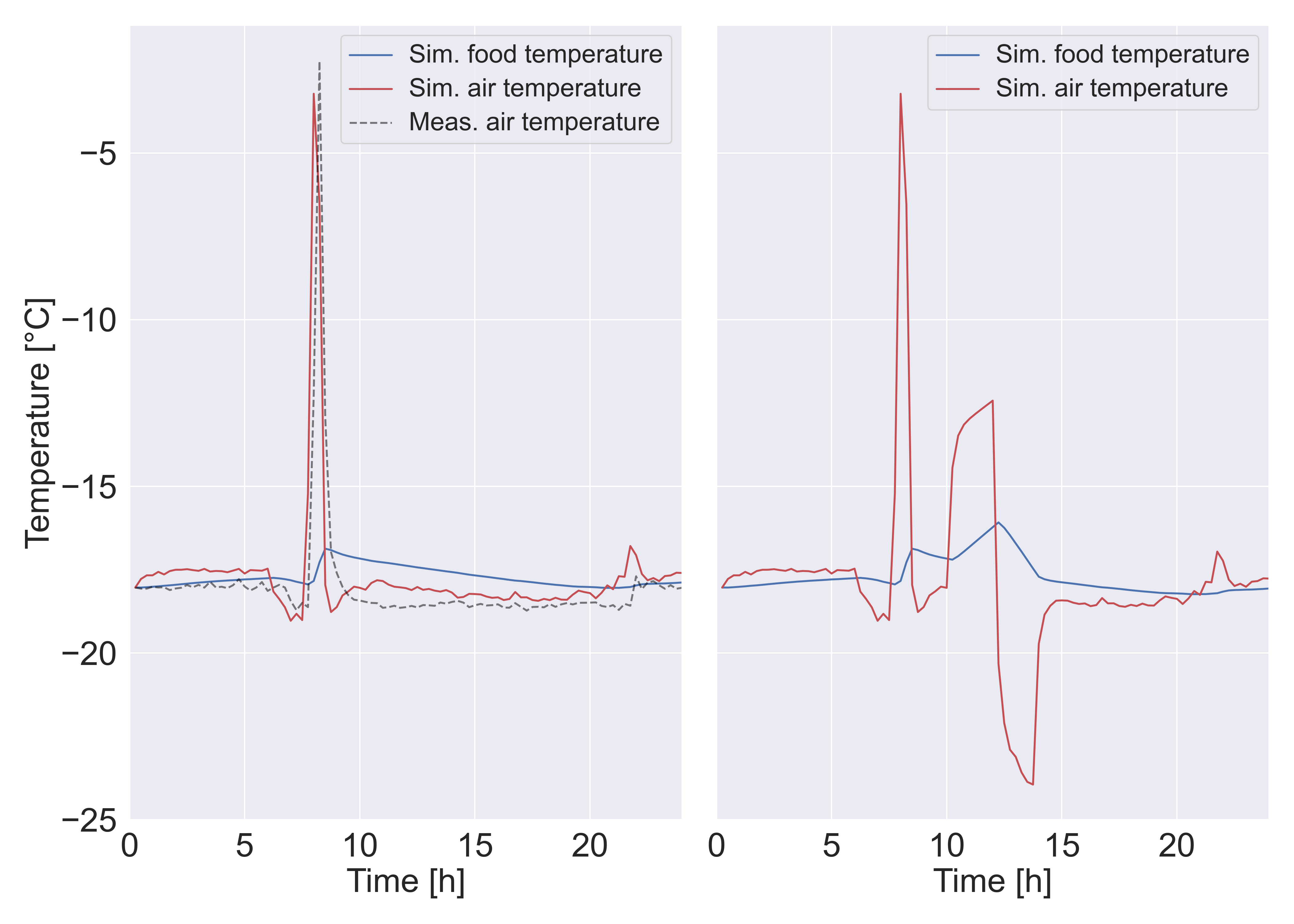}
    \caption{ \textbf{Left}: Simulation of food and air temperatures using (\ref{eq:2ndFreezerStateSpace}) with parameters in Table \ref{tab:parameter_estimates}, and its comparison to the measured air temperature. \textbf{Right}: Simulation where power is turned off for two hours with a subsequent rebound at the nominal power until the food temperature is back to its normal value.}
    \label{fig:2ndFreezerModelSimulation}
\end{figure}

Furthermore, Fig. \ref{fig:2ndFreezerModelSimulation} (left) shows a 24-hour simulation of  (\ref{eq:2ndFreezerStateSpace}). It is observed that the simulation is  reasonable and closely follows the measured air temperature, although a slight bias is observed as the simulated temperature is a bit higher. Such a visual validation is important because the model will be embedded later in an optimization model in Section \ref{sec:OptimizationModel}.

Ideally, the validation of (\ref{eq:2ndFreezerStateSpace}) would also include real measurements from the air and food temperature in a freezer during demand-response events. However, by adhering to the fundamental physics governing the temperature dynamics as shown, the model is trusted to be accurate during demand-response events too. It brings an intuitive interpretation which can be used to understand the impact on temperature during a demand-response event.
In Fig. \ref{fig:2ndFreezerModelSimulation} (right), such a simulated example of a demand-response event is shown. It can be observed how the air temperature increases when the power is turned off, and how it decreases when the power is turned back on. The food temperature is much more stable and only changes slightly, as expected. The \textit{rebound} occurs until the food temperature is back to its normal value.

%\noident
\textit{Notational remark}: Hereafter, $t \in \{1, \ldots, J=96\}$ represents the index for 15-minutes time steps, whereas $h\in \{1, \ldots, 24 \}$ denotes the index for hours.

\subsection{mFRR}\label{sec:mFRR}
\begin{figure}[t]
    \centering
    \definecolor{RectangleColor}{RGB}{197,224,180}
\definecolor{LightBlue}{RGB}{173, 216, 230}

\begin{tikzpicture}[scale=0.75,every text node part/.style={align=center}]

  % Legend box with LightBlue color
%\node[draw, rectangle, fill=LightBlue, rounded corners=3pt, minimum width=3cm, minimum height=0.7cm, align=center] at (13.5,6) {\textbf{Load shifting}};  
  % Legend box with RectangleColor
%\node[draw, rectangle,fill=RectangleColor, rounded corners=3pt, minimum width=3cm, minimum height=0.7cm, align=center] at (9,6) {\textbf{mFRR market}};

  \draw[rounded corners, fill=RectangleColor, very thick] (0, 3) rectangle (4, 5) node[midway]{\Large mFRR \\ \Large bidding};
  \node at (2,2) {\Large  $p_{h}^{\rm{r},\uparrow}$};
  \draw[dashed, black, line width=1.5pt] (4.5,-1) -- (4.5,5);
  \node at (4.5,-2) {\Large D$-$1};
  \node at (4.5,-2.7) {\large (9:30am)};
  \node[align=center, draw=none, fill=none, rotate=90] at (4.8,2.1) {mFRR market gate closure};

  \draw[dashed, black, line width=1.5pt] (7,-1) -- (7,5);
  \node at (7,-2) {\Large D$-$1};
  \node at (7,-2.7) {\large (12:00pm)};
  \node[align=center, draw=none, fill=none, rotate=90] at (6.7,2.5) {Day-ahead market gate closure};
  \draw[rounded corners, fill=LightBlue, very thick] (7.375, 1) rectangle (11.375, 3) node[midway]{\Large Load  \\ \Large shifting};
  \node at (9.375,0) {\Large  $p_{h}^{\text{Base}}$};

  \draw[rounded corners, fill=RectangleColor, very thick] (12, 3) rectangle (16, 5) node[midway]{\Large Regulating  \\ \Large power \\ \Large bidding};
  \node at (14,2) { \Large $\lambda_{h,\omega}^{\text{bid}}$};
  \draw[dashed, black, line width=1.5pt] (16.25,-1) -- (16.25,5);
  \node[align=center, draw=none, fill=none, rotate=90] at (16.65,3.2) {Deadline for regulating bid submission};
  \node at (16.25,-2) {\Large D$-$1};
  \node at (16.25,-2.7) {\large (5:00pm)};

  \draw[rounded corners, fill=RectangleColor, very thick] (17.5, 3) rectangle (21.5, 5) node[midway]{\Large mFRR \\ \Large activation};
  \node at (19.5,2) {\Large $\Gamma_{h,\omega}$};
  % {\Large $\Bigl( \bm{p}^{\rm{r},\uparrow}, \bm{\lambda}_{\omega}^{\text{bid}}, \bm{p}_{\omega}^{b,\uparrow}, \bm{p}_{\omega}^{b,\downarrow}, \bm{s}_{\omega}, \bm{T}_{\omega}^{c}, \bm{T}_{\omega}^{f}, \notag \\ \quad & \quad \bm{T}_{\omega}^{c, B}, \bm{T}_{\omega}^{f,B}, \bm{\phi}_{\omega}, \bm{g}_{\omega} \Bigr)$};
  \node at (19.5,-2) {\Large D};
  \node at (19.5,-2.7) {\large (12pm -- 12am)};

  % Right arrow
  \node (source) at (0, -1) {};
  \node (destination) at (24,-1){};
  \draw[->, line width=0.65mm, black](source)--(destination);

\end{tikzpicture}
    \caption{\textcolor{black}{Decision-making timeline and variables for the TCL when providing flexibility either in the form of mFRR services (green boxes) or load shifting (blue box).}}
    \label{fig:timeline_mfrr_variables}
\end{figure}
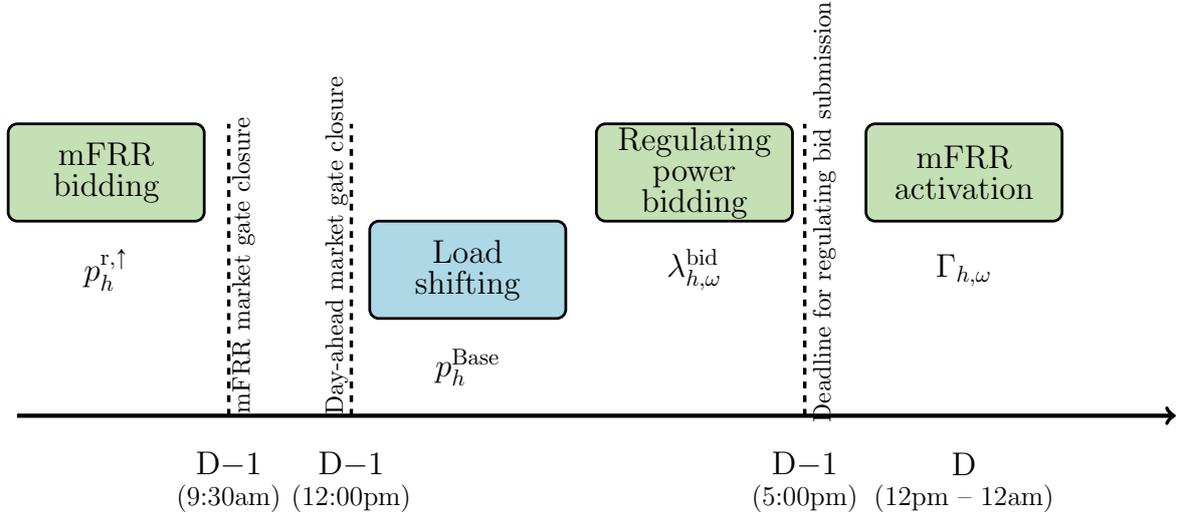

\textcolor{black}{The green boxes in Fig.\ref{fig:timeline_mfrr_variables} show the decision making timeline for the TCL to participate in the Danish mFRR market. The mFRR market in Denmark only trades up-regulation services.} One should note  that BRPs can choose \textit{not} to bid in the reserve market and  only bid in the real time for up- and down-regulation. We consider a case wherein the TCL delivers reservation through a BRP, since the payment is received for both reservation and activation. All prices considered are for the bidding zone DK2 (eastern Denmark).

% \begin{figure}[!t]
%     \centering
%     \includestandalone[width=\columnwidth]{tex/figures/timeline_mfrr_tikz}
%     \caption{Timeline of the Danish mFRR market. Symbols $\bm{p}$ denote power (in MW), whereas symbols  $\bm{\lambda}$ refer to price (in DKK/MW).}
%     \label{fig:timeline_mfrr}
% \end{figure}

According to green boxes in Fig. \ref{fig:timeline_mfrr_variables}, at 9:30am of day $\rm{D}$-1, the BRP submits reserve capacity bids $p_{h}^{\rm{r},\uparrow}$ $\forall{h}$ to the mFRR market for the next day $\rm{D}$. If accepted, she is paid at the reservation price $\lambda_{h}^{\rm{r}}$. Therefore, she earns $\lambda_{h}^{\rm{r}}p_{h}^{\rm{r},\uparrow}$. This happens \textit{before} the day-ahead (spot) market clearing at noon for which the BRP buys energy for her expected demand $P_{h}^{\rm{Base}}$ at the spot price $\lambda_{h}^{\rm{s}}$, \textcolor{black}{(which is not a decision variable in our model)}. After that, until 5pm, a regulating power bid, $\lambda_{h}^{\rm{bid}}$, must be submitted by the BRP for each hour of day $\rm{D}$, where $p_{h}^{\rm{r},\uparrow} > 0$ \cite{energinet:Systemydelser}. In hour $h$ of day $\rm{D}$, the reserves are activated if the following three conditions hold: (\textit{i}) there is a reservation accepted, i.e., $p_{h}^{\rm{r},\uparrow} > 0$, (\textit{ii}) the balancing price is higher than the spot price, i.e., $\lambda_{h}^{\rm{b}} > \lambda_{h}^{\rm{s}}$, and the regulating power bid is lower than or equal to the balancing price minus the spot price, i.e., $\lambda_{h}^{\rm{bid}} \leq  \lambda_{h}^{\rm{b}} - \lambda_{h}^{\rm{s}}$. If all these three conditions are met, the BRP is paid at the balancing price $\lambda_{h}^{\rm{b}}$ times her actual up-regulation $p_{h}^{\rm{b},\uparrow}$. The BRP may  incur a cost due to any subsequent rebound $p_{h}^{\rm{b},\downarrow}$. Furthermore, the BRP should pay at the penalty price $\lambda^{\rm{p}}$ for $s_{h} = \text{max}\{0, p_{h}^{\rm{r},\uparrow} - p_{h}^{\rm{b},\uparrow}$\}, i.e., if she fails delivering her promised reserve. In reality, $p_{h}^{\rm{b},\uparrow}$ is determined by the TSO, but for simplicity, we assume a bid is always activated in its entirety.
Accordingly, the objective function of the BRP maximizing the flexibility value over a day (via mFRR provision) is
\begin{align}\label{eq:mFRRObjective}
    \underbrace{\sum_{h=1}^{24}\lambda_{h}^{\rm{r}} p^{\rm{r}, \uparrow}_{h}}_{\textrm{Reservation payment}} + \underbrace{\sum_{h=1}^{24}  \lambda_{h}^{\rm{b}} p^{\rm{b},\uparrow}_{h}}_{\textrm{Activation payment}} - \underbrace{\sum_{h=1}^{24}  \lambda_{h}^{\rm{b}} p^{\rm{b},\downarrow}_{h}}_{\textrm{Rebound cost}} - \underbrace{ \sum_{h=1}^{24}  \lambda^{\rm{p}}s_{h}}_{\textrm{Penalty cost}}.
\end{align}

\textcolor{black}{Note that an additional term containing the energy cost $\sum_{h=1}^{24} \lambda^{\rm{s}}_{h}P^{\rm{Base}}_{h}$ is removed from \eqref{eq:mFRRObjective}, since it is a fixed term in our setup. This term contains $\lambda^{\rm{s}}_{h}$ and $P^{\rm{Base}}_{h}$, which are both parameters, and therefore, can be omitted from the objective function. However, we will consider it when numerically comparing cost savings associated with two alternatives of mFRR provision and load shifting.}

\subsection{Load shifting}
Another option for utilizing flexibility is to shift the load to a different time according to the spot market prices which are known already 12-36 hours in advance. \textcolor{black}{For this, the TCL makes a load shifting decision right after the day-ahead market as shown by the blue box in Fig. \ref{fig:timeline_mfrr_variables}}. \textcolor{black}{Note that in Denmark and generally in Europe, day-ahead prices are directly exposed to consumers with no fixed time-of-use price in between as is often the practice in North America}. %Then, it is simply a matter of consuming in low-price hours.
For a TCL, there are additional constraints to how energy can be shifted due to the rebound effect. First, there can be temperature constraints which will result in less energy being shifted. Second, the rebound must happen immediately after reducing power consumption. Otherwise, the temperature deviation becomes too big for too long.
The cost saving from load shifting is directly proportional to the volume and price difference between the baseline load and shifted load as given by
\begin{equation}\label{eq:load_shifting_savings}
    \sum_{h=1}^{24} \lambda^{\rm{s}}_{h} \big(p^{\rm{Base}}_{h} -  p_{h}\big),
\end{equation}
where $p_{h}$ is the power profile when the load is shifted, whereas $p^{\rm{Base}}_{h}$ is the baseline in the day-ahead stage.

Since the load shifting action only occurs \textit{after} the day-ahead market clearing (cf. Fig. \ref{fig:timeline_mfrr_variables}), the BRP has already bought $\lambda^{\rm{s}}_{h} p^{\rm{Base}}_{h}$ and any deviation results in an imbalance for the BRP. In this work, we look at the case where the flexible demand acts selfishly and excludes the BRP from her load shifting action. Therefore, the objective function for \textcolor{black}{the underlying TCL} simply reduces to $\sum_{h=1}^{24} \lambda_{h}^{\rm{s}} p_{h}$.

\section{Optimization model and solution strategy}\label{sec:OptimizationModel}

This section presents the optimization problem. The rest of this section is organized as follows. First, the time sequence of decisions for participation in the mFRR market is presented. Second, we explain how scenarios for price data are generated. Third, we present the model formulation. Fourth, we discuss how the bidding policy is implemented. Lastly, we show how a scenario decomposition method with an ADMM strategy is used to solve the optimization model.

\subsection{Time sequence for decision making}
Fig. \ref{fig:timeline_mfrr_variables} shows the stages for making decisions in the mFRR market \textcolor{black}{and for load shifting}. In the first stage in day $\rm{D}$-1, the BRP makes a reservation bid decision $p_{h}^{\rm{r},\uparrow}$ for every hour $h$ of day $\rm{D}$,  while being uncertain about input parameters in the next stages, including day-ahead market prices $ \lambda_{h}^{\rm{s}}$  and balancing market prices $ \lambda_{h}^{\rm{b}}$. We assume the BRP is not uncertain about mFRR market prices $ \lambda_{h}^{\rm{r}}$.

For simplicity and \textcolor{black}{to} avoid the need for developing a multi-stage stochastic program, we merge the second, third, and fourth stages in Fig. \ref{fig:timeline_mfrr_variables} as one, and call it the second stage. By this, $p_{h}^{\rm{r},\uparrow}$ is the first-stage variable, whereas the second stage variables, indexed by scenario $\omega$, include
the regulating power bid $\lambda_{h,\omega}^{\rm{bid}}$ and the set of real-time variables $\Gamma_{h,\omega} = \{ p_{h,\omega}, p_{h,\omega}^{\rm{b},\uparrow}$, $p_{h,\omega}^{\rm{b},\downarrow}$, $s_{h,\omega}$, $T_{h,\omega}^{\rm{c}}$, $T_{h,\omega}^{\rm{f}}$, $T_{h,\omega}^{\rm{c,\text{Base}}}$, $T_{h,\omega}^{\rm{f,\text{Base}}}$, $\phi_{h,\omega}$, $g_{h,\omega}, u^{\uparrow}_{h,\omega}, u^{\downarrow}_{h,\omega}, y^{\uparrow}_{h,\omega}, y^{\downarrow}_{h,\omega}, z^{\uparrow}_{h,\omega}, z^{\downarrow}_{h,\omega} \}$. This set contains the real-time power activated, as well as auxiliary variables for identifying up- and down-regulation,  temperature dynamics, and when to deliver up-regulation according to the bid and prices. See the Appendix for a detailed description. Note that down-regulation refers to the rebound effect of the freezer.
% \begin{figure}[!t]\label{fig:timeline_mfrr_variables}
%     \centering
%     \includegraphics[width=\columnwidth]{tex/figures/timeline_mfrr_variables.png}
%     \caption{Variables related to mFRR up-regulation decisions. The asterisk indicates the first-stage decision.}
% \end{figure}

\subsection{Scenario generation}\label{sec:scenario_generation}
To make decisions on reservation capacity $p_{h}^{\rm{r},\uparrow}$ and regulating power bid $\lambda_{h,\omega}^{\rm{bid}}$, we generate a set of scenarios $\omega \in \Omega$ for spot and balancing price data., i.e., $ \lambda_{h,\omega}^{\rm{s}}$ and $ \lambda_{h,\omega}^{\rm{b}}$. Each scenario contains spot and balancing price data for the entire day in question. The number of scenarios is $|\Omega|$, and we assume all scenarios are equiprobable. We refer to them as \textit{in-sample} scenarios. We use two different strategies for in-sample scenario generation: (\textit{i}) considering historical spot and balancing prices in DK2 in 2021.
%, with different cases where the number of scenarios $|\Omega|$ is 1, 5, 10, 20, 30, 40, 50, 100, and 250. 
(\textit{ii}) considering prices of the most recent five days (lookback strategy). \textcolor{black}{This first scenario generation strategy represents the case where an aggregator relies on a simple, fixed policy without the need to run a daily optimization as opposed to the second  strategy, although it might exhibit a better performance in case the underlying uncertainty is non-stationary. The first  strategy uses 50 scenarios. Using a sensitivity analysis, we found out that increasing the number of scenarios beyond 50 does not improve the performance remarkably as explained in Appendix B. We made a similar observation for choosing five days in the second solution strategy.}
In both scenario generation strategies, balancing prices $\lambda_{h,\omega}^{\rm{b}}$ are sampled in the following way: First, an integer $v$ is sampled uniformly from $\{0, \ldots, 24\}$ which represents the total number of up-regulation hours in a day. We then sample spot and balancing price differentials from a single day within the set of all days where up-regulation happened $v$ times. This constitutes one scenario and is repeated $|\Omega|$ times. In this way, days where up-regulation happened are essentially up-sampled, and the model learns more when up-regulation happens than otherwise. The in-sample results are systematically compared against the same set of unseen \textit{Out-of-Sample} (OOS) scenarios, which are DK2 prices for 2022.

For load shifting, the solution approach is simply to solve a deterministic optimization problem for the next day, since the day-ahead (spot) market clearing happens in advance, and thereby the spot market prices are known.

\subsection{Model formulation}
Recall that \eqref{eq:mFRRObjective} gives the objective function of a BRP participating in the mFRR market, but in a deterministic setup.  The optimization problem of such a BRP in a two-stage stochastic programming format is
\begin{subequations}\label{P1:compact_model}
    \begin{align}
        \underset{\bm{p}^{\rm{r},\uparrow}, \bm{\lambda}_{\omega}^{\rm{bid}}, \bm{\Gamma}_{\omega}}{\text{Maximize}} \  & f(\bm{p}^{\rm{r},\uparrow}) + \sum_{\omega \in \Omega} \pi_{\omega} \ g(\bm{\Gamma}_{\omega}) \label{P1:eq1}
        \\
        \text{s.t.} \                                                                                                   & h(\bm{p}^{\rm{r},\uparrow}, \bm{\lambda}_{\omega}^{\rm{bid}}, \bm{\Gamma}_{\omega}) \leq 0, \ \forall{\omega}\label{P1:eq2}                                                                                                 \\
        \                                                                                                               & \text{State-space model } (\ref{eq:2ndFreezerStateSpace}), \ \forall{\omega} \label{P1:eq3}
        \\
        \                                                                                                               & \bm{T}_{\omega}^{\rm{c}}, \bm{T}_{\omega}^{\rm{f}},  \bm{T}_{\omega}^{\rm{c,\text{Base}}}, \bm{T}_{\omega}^{\rm{f,\text{Base}}}\in \mathbb{R}  \label{P1:eq4}
        \\
        \                                                                                                               & \bm{p}_{\omega}, \bm{p}^{\rm{r},\uparrow}, \bm{\lambda}_{\omega}^{\rm{bid}}, \bm{p}_{\omega}^{\rm{b},\uparrow}, \bm{p}_{\omega}^{\rm{b},\downarrow}, \bm{s}_{\omega},  \bm{\phi}_{\omega}\in \mathbb{R}^{+}  \label{P1:eq5}
        \\
        \                                                                                                               & \bm{g}_{\omega}, \bm{u}^{\uparrow}_{\omega}, \bm{z}^{\uparrow}_{\omega}, \bm{y}^{\uparrow}_{\omega}, \bm{u}^{\downarrow}_{\omega}, \bm{z}^{\downarrow}_{\omega}, \bm{y}^{\downarrow}_{\omega} \in \{0,1\},  \label{P1:eq6}
    \end{align}
\end{subequations}
where bold symbols represent vectors. For example, the vector $\bm{p}^{\rm{r},\uparrow}$ includes $p_{h}^{\rm{r},\uparrow} \ \forall{h}$. Parameter $\pi_{\omega}$ is the probability assigned to in-sample scenario $\omega$. Here, we present optimization \eqref{P1:compact_model} in a compact form using functions $f(.)$, $g(.)$, and $h(.)$. The detailed formulation is given in the Appendix. Constraint (\ref{P1:eq2}) includes all power and activation related limits, whereas (\ref{P1:eq3}) models temperature dynamics in the freezer. Constraints (\ref{P1:eq4})-(\ref{P1:eq5}) declare continuous and binary variables. Note that $\mathbb{R}$ and $\mathbb{R}^{+}$ denote free and non-negative real numbers, respectively. The optimization model \eqref{P1:compact_model} is a MILP problem.

For optimal decision making for load shifting, \eqref{P1:compact_model} is simplified by removing scenarios as well as reservation and bid constraints, and replacing the objective function (\ref{P1:eq1}) by minimizing the total power purchase cost $\bm{\lambda}^{\rm{s}} \bm{p}$.

\subsection{Regulating power bidding implementation}\label{sec:mFRR_bidding_implementation}
Recall from  Fig. \ref{fig:timeline_mfrr_variables} that we have merged three stages, by which the regulating power bidding decisions $\lambda_{h,\omega}^{\rm{bid}}$ and real-time operational decisions $\Gamma_{h,\omega}$ become second-stage variables. This is the reason both set of variables $\lambda_{h,\omega}^{\rm{bid}}$ and $\Gamma_{h,\omega}$ are similarly indexed by $\omega$, while in reality, the decision $\lambda_{h,\omega}^{\rm{bid}}$ should be made before $\Gamma_{h,\omega}$. This also challenges the ex-post OOS simulation. To resolve it, we use a learning policy, such that we replace the scenario-indexed variable $\lambda_{h,\omega}^{\rm{bid}}$ in \eqref{P1:compact_model} by $\alpha q(\lambda_{h,\omega}^{\rm{s}}) + \beta$, where $q(.)$ is an arbitrarily selected function. In addition, $\alpha$ and $\beta$ are non-negative first-stage  variables (they are not indexed by $\omega$). This replacement shrinks the degree of freedom for the BRP compared to \eqref{P1:compact_model}, but makes it more practical to be used. The reason is that, by this trick, the regulating power bidding decision to be made at 5pm of day $\rm{D}$-1 becomes a first-stage decision, dependent not only on policies $\alpha$ and $\beta$, but also on uncertain spot prices $\lambda_{h,\omega}^{\rm{s}}$. In other words, by using the in-sample scenarios $\omega$, the BRP obtains optimal values for $\alpha$ and $\beta$ at 9:30am of day $\rm{D}$-1. Then, she waits to see the spot prices at noon of day $\rm{D}$-1, and  submits her regulating power bids at 5pm of day $\rm{D}$-1. In our simulations, we found out that the selection of $q(\lambda_{h,\omega}^{\rm{s}})$ as the difference of spot prices in subsequent hours works comparatively more satisfactory ex-post, as it suits better to accommodate the rebound effect. By this, the BRP prefers to be activated when the spot price during rebound (down-regulation) is comparatively lower. Therefore, we replace $\lambda_{h,\omega}^{\rm{bid}}$ in \eqref{P1:compact_model} by
\begin{align}\label{eq:affine_policy}
     & \alpha ( \lambda_{h+1,\omega}^{\rm{s}} - \lambda_{h,\omega}^{\rm{s}}) + \lambda_{h,\omega}^{\rm{s}} + \beta, \ \forall{\omega}, \forall{h} \in \{1, \ldots, 23\}.
\end{align}

%To solve Problem (\ref{P1:compact_model}), we first need to specify a bidding policy that can readily be used OOS. We do so by choosing an affine bidding policy. Afterwards, it is shown how the bidding policy is implemented using McCormick relaxation.

%\subsubsection{Affine bidding policy}

%A bidding policy needs to be easy to follow OOS for the trader. We choose an affine bidding policy, i.e., a linear function of the spot price. The bidding policy is given by:

%Variables $\alpha$ and $\beta$ are then learned IS and fixed for OOS evaluation. After the day-ahead market clearing, (\ref{eq:affine_policy}) can easily be used to specify bids for the next day.

Another implementation challenge is to enforce price conditions under which the mFRR reservation is activated.
Recall from Section \ref{sec:mFRR}, for the activation at hour $h$ under scenario $\omega$, it is necessary to hold $\lambda_{h,\omega}^{\rm{bid}} \leq  \lambda_{h,\omega}^{\rm{b}} - \lambda_{h,\omega}^{\rm{s}}$ and $ \lambda_{h,\omega}^{\rm{b}} > \lambda_{h,\omega}^{\rm{s}}$. These conditions can be equivalently enforced as
%
%activation of mFRR reservation only happens when certain price conditions are met. This is formalized in the following constraint:
%
\begin{equation}\label{eq:bid_constraint}
    p^{\rm{b}, \uparrow}_{h,\omega} + s_{h,\omega} \geq p^{\rm{r},\uparrow}_{h}  \mathbbm{1}^{\big(\lambda_{h,\omega}^{\rm{bid}} \leq  \lambda_{h,\omega}^{\rm{b}} - \lambda_{h,\omega}^{\rm{s}} \ \text{and} \ \lambda^{\rm{b}}_{h,\omega} > \lambda^{\rm{s}}_{h,\omega}\big)},
\end{equation}
where $\mathbbm{1}^{(.)}$ is 1 if conditions (.) are met, otherwise it is zero. If the non-negative slack variable $s_{h,\omega}$ takes a non-zero value, it shows that the BRP fails in the activation stage, and therefore will be penalized. The challenge is that \eqref{eq:bid_constraint} makes a condition on variable $\lambda_{h,\omega}^{\rm{bid}}$, or equivalently on $\alpha$ and $\beta$ as defined in \eqref{eq:affine_policy}. This makes \eqref{eq:bid_constraint} non-linear. To linearize it, we use the McCormick relaxation technique \cite{mccormick1976computability} and define auxiliary  variables $\phi_{h,\omega} \in \mathbb{R}^{+}$ and $g_{h,\omega} \in \{0, 1\}$. By this, we replace \eqref{eq:bid_constraint} for every hour $h$ and scenario $\omega$ by a set of mixed-integer linear constraints as
%
%Eq. (\ref{eq:bid_constraint}) shows how real-time up-regulation plus a slack variable must be greater than or equal to the reservation if the bid is lower than the balancing price and if up-regulation is needed in hour $h$. It is a bi-linear constraint so McCormick relaxation \cite{mccormick1976computability} is used to convert (\ref{eq:bid_constraint}) to a linear constraint by introducing auxiliary variables, $\phi_{h,\omega}$ and $g_{h,\omega}$:
%
\begin{subequations}\label{eq:bid_constraint_relaxed}
    \begin{align}
         & \lambda_{h,\omega}^{\rm{bid}} - M  (1 - g_{h,\omega}) \leq \lambda_{h,\omega}^{\rm{b}} - \lambda_{h,\omega}^{\rm{s}} \leq \lambda_{h,\omega}^{\rm{bid}} + M  g_{h,\omega},                               \label{con_bid:subeq1} \\
         & p^{\rm{b}, \uparrow}_{h,\omega} \leq \phi_{h,\omega}  \mathbbm{1}^{\lambda^{\rm{b}}_{h,\omega} > \lambda^{\rm{s}}_{h,\omega}}, \                          \label{con_bid:subeq3}                                                \\
         & p^{\rm{b}, \uparrow}_{h,\omega} + s_{h,\omega} \geq \phi_{h,\omega}  \mathbbm{1}^{\lambda^{\rm{b}}_{h,\omega} > \lambda^{\rm{s}}_{h,\omega}}, \            \label{con_bid:subeq4}                                               \\
         & -g_{h,\omega}  M \leq \phi_{h,\omega} \leq g_{h,\omega}  M,                                   \label{con_bid:subeq5}                                                                                                            \\
         & -(1 - g_{h,\omega})  M \leq \phi_{h,\omega} - p^{\rm{r},\uparrow}_{h} \leq (1 - g_{h,\omega}) M, \                                                                                    \label{con_bid:subeq7}  \
        % \lambda_{h,\omega}^{\rm{bid}} \leq \lambda^{Max} \label{con_bid:subeq9}
    \end{align}
\end{subequations}
where $M$ is a large enough positive constant.
Constraint \eqref{con_bid:subeq1} ensures that $g_{h,\omega} = 1$ when $\lambda_{h,\omega}^{\rm{b}} - \lambda_{h,\omega}^{\rm{s}} \geq \lambda^{\rm{bid}}_{h, \omega}$. Otherwise, it sets $g_{h,\omega} = 0$. Constraints \eqref{con_bid:subeq3}-\eqref{con_bid:subeq4} set the up-regulation equal to $\phi_{h,\omega}$ (or incurs a penalty through $s_{h,\omega}$) if there is an up-regulation event in the system, i.e., if $\mathbbm{1}^{\lambda^{\rm{b}}_{h,\omega} > \lambda^{\rm{s}}_{h,\omega}} = 1$. Constraint \eqref{con_bid:subeq5} enforces  $\phi_{h,\omega} = 0$ when $g_{h,\omega} = 0$, implying the balancing price minus the spot price is smaller than the power regulating bid. Constraint \eqref{con_bid:subeq7} ensures that $\phi_{h,\omega}$ is equal to the reservation capacity $p^{\rm{r},\uparrow}_{h}$ whenever $g_{h,\omega} = 1$, i.e., if $\lambda_{h,\omega}^{\rm{b}} - \lambda_{h,\omega}^{\rm{s}} \geq \lambda^{\rm{bid}}_{h, \omega}$. Note that in the final model,  $\lambda^{\rm{bid}}_{h, \omega}$ in \eqref{eq:bid_constraint_relaxed} should be replaced as defined in \eqref{eq:affine_policy}. The resulting model formulation, which is a MILP, is given in the Appendix.

\subsection{Scenario decomposition with ADMM}\label{sec:admm}
By increasing the number of scenarios,  the proposed stochastic program  quickly becomes computationally intractable due to the number of binary variables and intertemporal constraints. To resolve it, we apply a scenario decomposition approach built upon the ADMM algorithm \cite{boyd2011distributed}. To do so, we relax non-anticipativity constraints by adding a scenario index to the first-stage variables, i.e.,
\begin{equation}\label{eq:non_anticipativity}
    p_{h}^{\rm{r},\uparrow} \rightarrow p^{\rm{r},\uparrow}_{h,\omega}, \ \alpha \rightarrow \alpha_{\omega}, \ \beta \rightarrow \beta_{\omega}.
\end{equation}

This decomposes the original problem to a set of deterministic sub-problems, one per scenario, to be solved in parallel. A quadratic regularizer is added to the objective function of every subproblem, making it a mixed-integer quadratic program.
The ADMM algorithm is iterative. The convergence happens when we achieve a consensus over sub-problems on first-stage  variables. Due to having binary variables in the original problem, this ADMM algorithm is eventually a heuristic \cite{hong2016convergence}, i.e., it may not converge to optimality. However, we have observed in a case with a limited number of scenarios for which we can also solve the original MILP problem directly, the proposed ADMM exhibits a satisfactory performance.

\section{Numerical results and discussion}\label{sec:results}

We consider five models to calculate the operational (energy) cost of a single freezer, including three models where the freezer provides mFRR services, one model where the freezer shifts the load in response to spot prices, and the last model, the so-called \textit{base cost}, where the freezer neither shifts the load nor provides mFRR services. The data availability for these five models is summarized in
Table \ref{tab:price_visibility}. All source codes and relevant data are publicly shared in \cite{code}.

\subsection{Load shifting vs mFRR: Which one is more appealing?}
\begin{table}[t]
    \caption{Data availability for each model.}
    \label{tab:price_visibility}
    \centering
    \begin{tabular}{llccc}
        \toprule
        Model         & Curve(s) in Fig. \ref{fig:cumulative_cost_comparison} & $\bm{\lambda}^{\rm{r},\uparrow}$ & $\bm{\lambda}^{\rm{s}}$ & $\bm{\lambda}^{\rm{b}}$ \\
        \midrule
        mFRR oracle   & Dashed black                                          & $\checkmark$                     & $\checkmark$            & $\checkmark$            \\
        mFRR cases    & Blue and red                                          & $\checkmark$                     & $\ast$                  & $\ast$                  \\
        Load shifting & Yellow                                                & N/A                              & $\checkmark$            & N/A                     \\
        Base cost     & Solid black                                           & N/A                              & $\checkmark$            & N/A                     \\
        \bottomrule
        \multicolumn{5}{l}{$\checkmark$: True (realized) data are available.}                                                                                        \\
        \multicolumn{5}{l}{$\ast$: Forecast data (in form of scenarios) are available.}                                                                              \\
        \multicolumn{5}{l}{N/A: Not applicable.}
    \end{tabular}
\end{table}

\begin{figure}[t]
    \centering
    \includegraphics[width=\columnwidth]{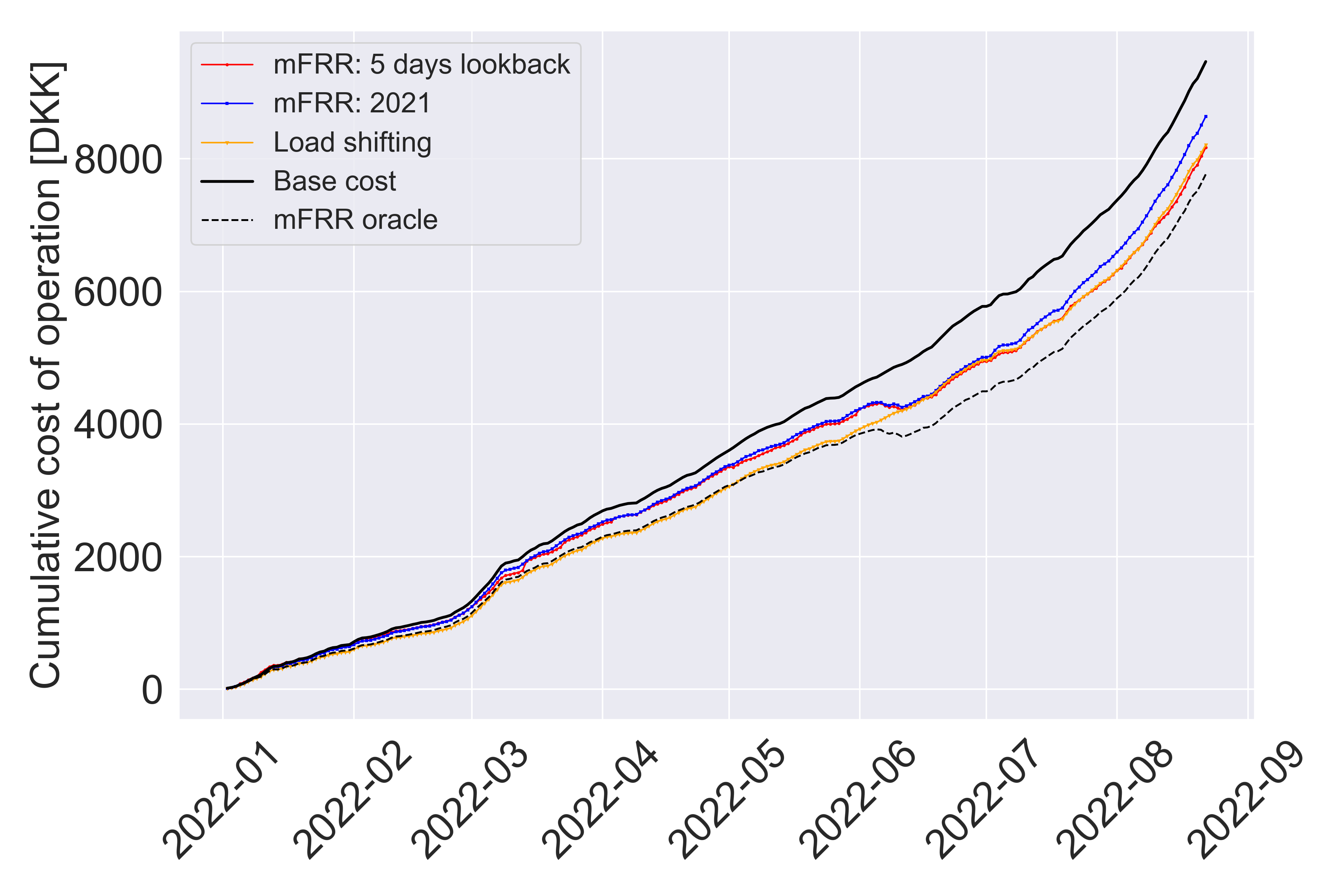}
    \caption{Out-of-sample cumulative operational cost for the freezer during the first nine months of 2022.}
    \label{fig:cumulative_cost_comparison}
\end{figure}

For all five models, Fig. \ref{fig:cumulative_cost_comparison} shows the out-of-sample cumulative operational cost of the single freezer during the first nine months of 2022. Recall that all these models have been compared fairly by an out-of-sample simulation against identical scenarios, i.e., real spot and balancing market prices from 2022.
The base cost (solid black) has only access to spot price data, and leads to the highest cost in Fig. \ref{fig:cumulative_cost_comparison}.
Blue and red curves (mFRR cases), both below the base cost curve, correspond to the cases where the freezer provides mFRR services, and uses scenarios to model spot and balancing market price uncertainties. Their difference comes from in-sample scenarios used: while the red curve uses five equiprobable scenarios coming from the most recent historical data (lookback strategy), the blue curve uses 50 equiprobable scenarios coming from 2021. The yellow curve shows the reduced cost of the freezer due to  load shifting. Finally, the dotted black curve (lowest in Fig. \ref{fig:cumulative_cost_comparison}) provides an oracle (ideal benchmark), where the freezer perfectly knows the spot and balancing market prices.

An interesting observation is that, in comparison to the base cost, both load shifting (yellow curve) and mFRR (red curve) bring a 13.9\% cost reduction, whereas the mFRR provision using 2021 data reduces the cost by  11.6\% (blue curve). Note that the cost savings reported for red and blue curves are not far  from the mFRR oracle, showing that the scenarios used are sufficiently adequate.
While this observation might be appealing to the freezer as load shifting requires a comparatively simpler decision-making process than the mFRR provision, it is not necessarily a desirable outcome for the power system. The mFRR  provision helps the system keep the supply-demand balance, while load shifting as a response to spot prices is not necessarily helping the system, and even in the worse case, the spot prices fixed before load shifting may no longer represent the supply-demand equilibrium. This calls system operators and regulators for potential changes, e.g., market redesign, to make the mFRR  provision more attractable.

In the case of load shifting, it is worth mentioning that the settlement cost of the BRP is ignored --- this reflects the self-interested \textcolor{black}{TCL}s.  While the \textcolor{black}{TCL} reduces her cost by load shifting, the corresponding BRP may incur a settlement cost due to the resulting imbalance between the true consumption and the day-ahead schedule. In practice, the BRP would have to pay for such an imbalance or at least buy/sell energy in the intra-day market to adjust her day-ahead schedule and consider potential load shifting actions by \textcolor{black}{TCL}s. However, loads are not necessarily aware of those costs for the BRP. Hence, they may have a strong incentive to exploit their flexibility for load shifting. Some larger flexible consumers, such as industrial and commercial loads, might not be exposed fully to spot prices, and for those consumers, load shifting could be less profitable. However, they may still have an incentive to change their deal with the BRP to get a full exposure.

In the case of mFRR  provision, it is   assumed that the entire revenue from reservation and activation go to the \textcolor{black}{TCL}. This neglects the fact that, in practice, the BRP requires a share of the revenue. Furthermore, there might be an aggregator or a technology provider who facilitates the aggregation and communication of the flexibility. In such a case, they would also request a share of the revenue. This can potentially reduce the revenue of \textcolor{black}{TCL}s by the mFRR provision.

\subsection{Technical results}
\begin{figure}[H]
    \centering
    \includegraphics[width=\columnwidth]{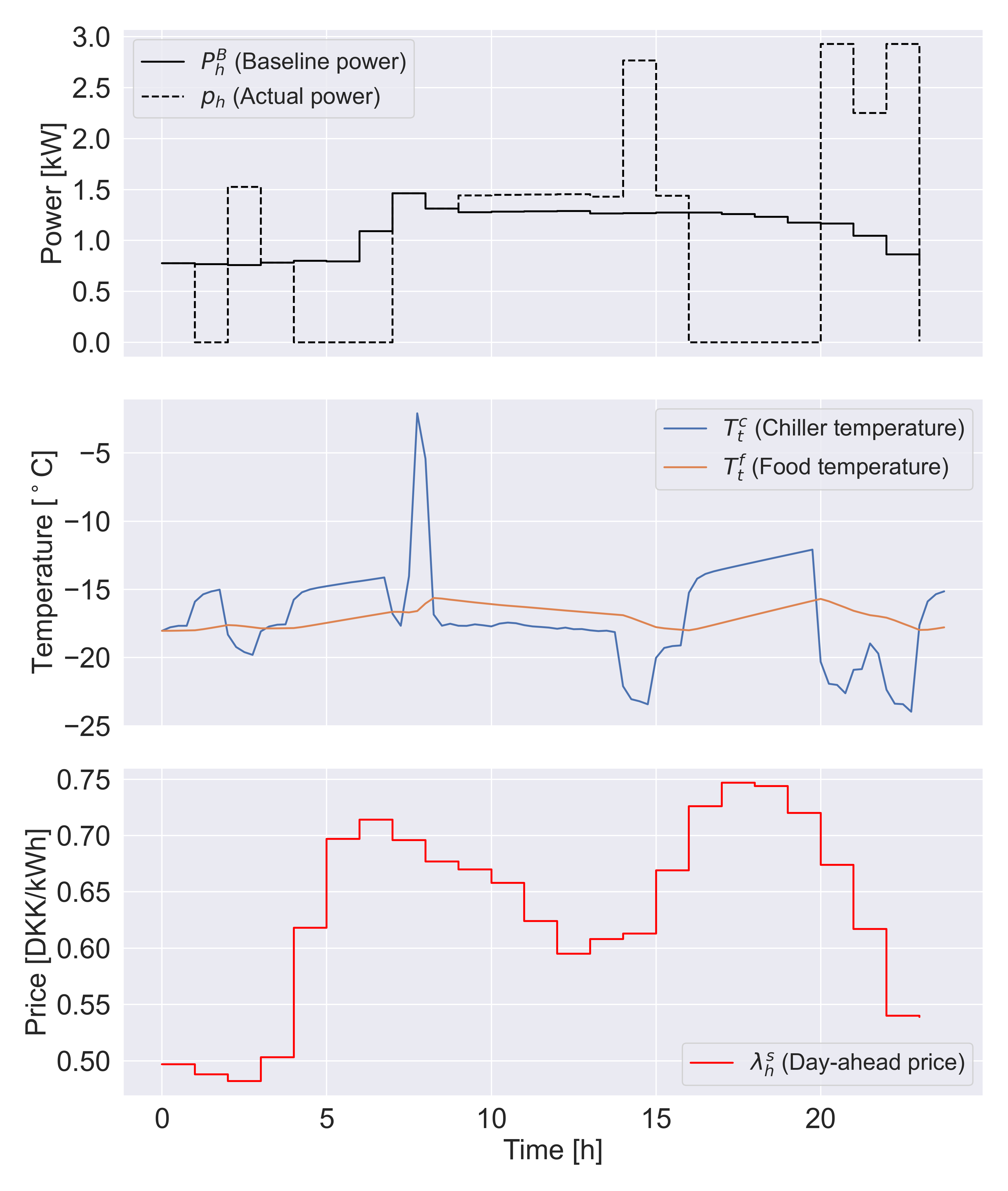}
    \caption{Example of load shifting in a representative day (an in-sample scenario). \textbf{Top}: Baseline consumption of the freezer and the power profile after load shifting. \textbf{Middle}: Air and food temperature dynamics. \textbf{Bottom}: Spot market prices.}
    \label{fig:fig_first_case}
\end{figure}

Fig. \ref{fig:fig_first_case} shows technical results for the case of load shifting in a representative day. For given spot prices (bottom plot), it is evident that the freezer shifts the load to low-price hours, especially in the later hours of the day (top plot). However, this has a remarkable effect on the air and food temperatures with large deviations from the normal set-point (middle plot), which in turn can have an impact on food degradation.

%, load shifting and mFRR are compared for the same day. For load shifting, it can clearly be seen how load is shifted to low-price hours, especially at the end of the day. But it also has a very significant effect on the temperature with large deviations from its normal setpoint. 

\begin{figure}[H]
    \centering
    \includegraphics[width=\columnwidth]{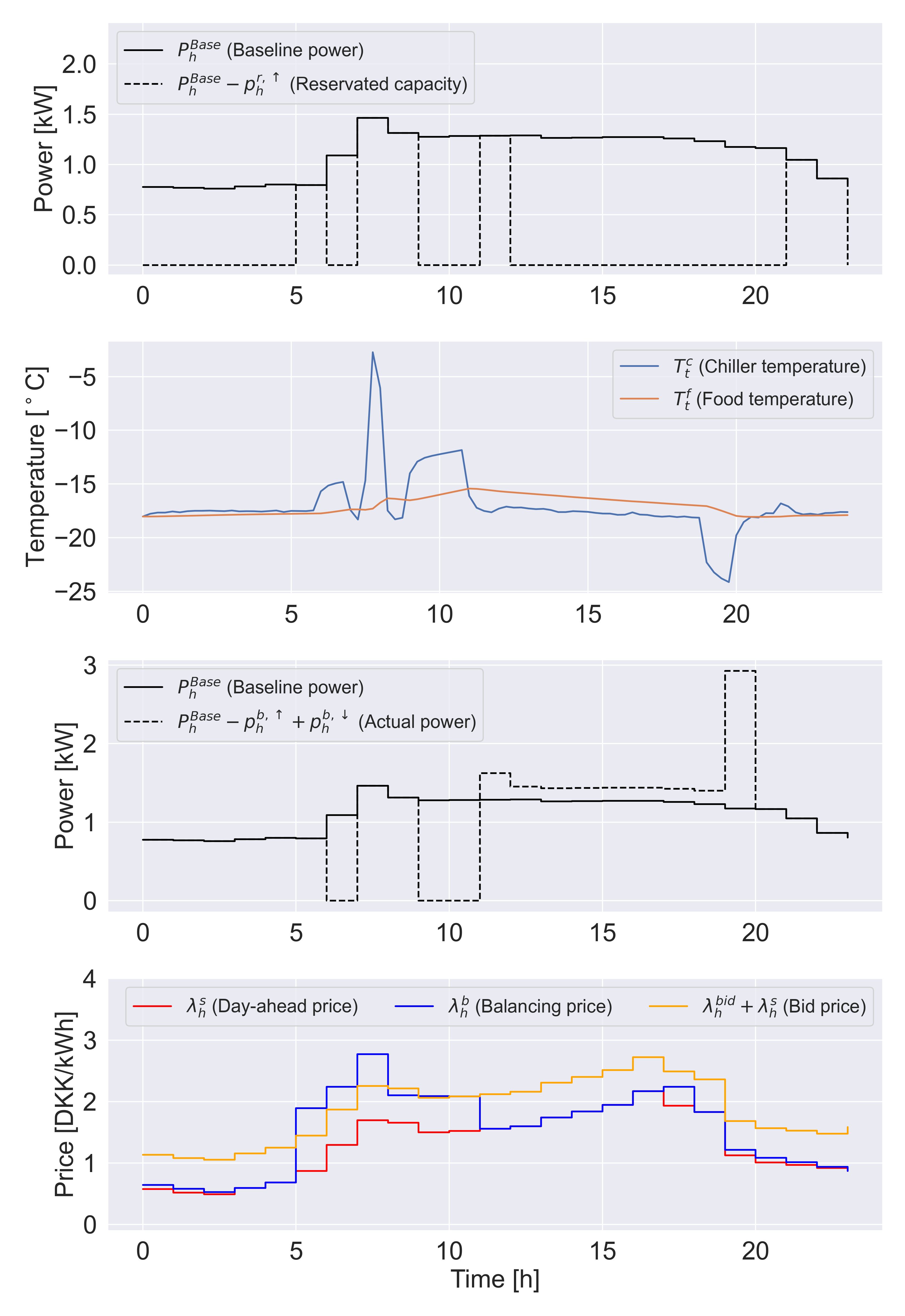}
    \caption{Example of mFRR  provision  in a representative day (an in-sample scenario). \textbf{Top}: Reservation capacities and baseline power of the freezer. \textbf{Upper middle}: Air and food temperature dynamics. \textbf{Lower middle}: mFRR activation in this specific scenario, i.e., when $\lambda_{h}^{\rm{bid}} \leq \lambda_{h}^{\rm{b}}-\lambda_{h}^{\rm{s}}$, $\lambda_{h}^{\rm{b}} > \lambda_{h}^{\rm{s}}$, and $p^{\rm{r},\uparrow}_{h} > 0$. \textbf{Bottom}: Spot and balancing market price as well as regulating power bids in this specific scenario.}
    \label{fig:fig_second_case}
\end{figure}

Fig. \ref{fig:fig_second_case} presents technical results for the case of mFRR  provision  in a representative day, whose spot and balancing market prices are given in the lower plot. The top plot shows the baseline power profile $P^{\text{Base}}_{h}$ and the mFRR reservation $p^{\rm{r}, \uparrow}_{h}$ sold. For a better illustration, we have plotted $P^{\text{Base}}_{h}-p^{\rm{r}, \uparrow}_{h}$, i.e., the power profile of the freezer if the full activation happens during the entire day. The third  plot of Fig. \ref{fig:fig_second_case} shows, for this specific day (as an in-sample scenario), the freezer is activated three times, i.e., in hours 6, 9, and 10. In these three hours, all conditions $\lambda_{h,\omega}^{\rm{bid}} \leq  \lambda_{h,\omega}^{\rm{b}} - \lambda_{h,\omega}^{\rm{s}}$, $ \lambda_{h,\omega}^{\rm{b}} > \lambda_{h,\omega}^{\rm{s}}$, and $p^{\rm{r},\uparrow}_{h} > 0$ hold (see the bottom plot). The rebound  happens in hours 11 to 19. The majority of rebound occurs in hour 19, as the balancing price is comparatively low. Finally, the second plot shows that the food and air temperature deviations in the freezer due to mFRR  provision  are smoother in comparison to  load shifting.

%For mFRR, the reservation is almost full for all hours during the day. The activation of reservation occurs when the bid price is lower than the balancing price which in this particular scenario only happens for three hours. The effect on the temperature is therefore much smaller than for load shifting. The model is also able to rebound smartly in hour 19 to avoid high rebound costs.

Our second main observation is as follows: Although the cost saving is higher for load shifting in comparison to mFRR  provision, it is directly proportional to the energy shifted, and therefore, the temperature deviation in the freezer. This is not the case for mFRR, as the reservation is not always activated.

\section{Conclusion}\label{sec:conclusion}

To balance the power grid and ensure its stability in the future, there is a need to integrate industrial and commercial flexible demands into existing ancillary services where mFRR is the most energy-intensive one. To that end, it is crucial to estimate and model flexibility from flexible demands, especially thermostatically controlled loads.

We investigated how a supermarket freezer can provide flexibility for mFRR and load shifting in Denmark, and explored which one provides a greater monetary incentive for a \textcolor{black}{TCL}. To this end, we used actual data  from a Danish supermarket. This was done by developing a second-order grey-box model of the temperature dynamics in the freezer with the food temperature as a latent state. In the state-space form, the model was directly incorporated as constraints into a two-stage stochastic MILP problem, whose objective is to maximize the monetary value from the freezer's flexibility. Two scenario generation strategies were implemented: one with a five-day lookback strategy on the spot and balancing market prices in DK2, and the other one  based on price data for those markets in 2021. For mFRR, we used a linear policy, and then linearized the conditions for activation via the McCormick relaxation method. For computational ease, we used an ADMM-based scenario decomposition technique.  An out-of-sample evaluation was done on unseen 2022 price data. We observed that load shifting is more profitable, but has a greater impact on the air and food temperatures in the freezer as opposed to mFRR that depends on the system state and bid price for activation.

From a policy perspective, it is concerning if load shifting happens to be \textcolor{black}{almost as} profitable \textcolor{black}{as} mFRR provision. It might be of great interest for TSOs to ensure flexible demands have an incentive to deliver mFRR to ensure stability in the power grid and avoid new peaks and imbalances. Such incentives could manifest in green credits for providing emission free mFRR or an add-on to the mFRR reservation price. From the perspective of the flexible demand, periods of high volatility in spot prices can readily be used for load shifting although great care should be taken with respect to temperature deviations. Provision of mFRR is more attractive with respect to temperature deviations, but it is much more complicated to get started with mFRR whereas load shifting does not depend on other market participants, hence load shifting is a simple and profitable revenue stream for flexible demands.

We made a set of simplifications and assumptions, whose impacts need to be explored for the future work. The revenue share of BRP and aggregator was not considered. This may change our finding that load shifting is more financially appealing to a \textcolor{black}{TCL}. As mentioned earlier, a single freezer or supermarket must be part of a larger portfolio through an aggregator in order to participate in the mFRR market. Such an aggregated portfolio has some issues that are neglected here, such as the baseline estimation for verification of the demand response, allocation of profits within the portfolio, and an accurate capacity estimation of the whole portfolio that bids in the mFRR market. Furthermore, the European mFRR markets will change from a 60-minute resolution to a 15-minute market in the next few years \cite{MARI}. This makes it more feasible for TCLs to participate in the mFRR market, given their sensitivity to large temperature deviations, and the fact that TCLs can get a passive income when there is no up-regulation need in the power grid.

\section*{Credit author statement}
Peter Gade conceived and designed study, developed the code, and write the initial draft of the manuscript. All authors contributed considerably to developing the idea, thoroughly discussed the model, scenarios, numerical results, and policy insights, and also revised the final manuscript.

\section*{Acknowledgement}

The authors would also like to thank Coop and AK-Centralen for providing the freezer and power data from a supermarket.

\section*{Appendix \textcolor{black}{A}: MILP problem formulation}\label{appendix:A}

\begingroup
\allowdisplaybreaks
The two-stage stochastic MILP problem for the freezer to optimally bid in the mFRR market reads as
% First, the objective function is presented. Then all auxillary variables and constraints are presented. Third, constraints related to the power consumption for the freezer is shown. Fourth, the physical constraints for the temperatures are presented. Lastly, the rebound constraints are presented.
% \subsubsection{Objective function}\label{sec:objective_function}
%
\begin{subequations}\label{P2:FinalModel}
    \begin{align}
           & \underset{p_{h}^{\rm{r},\uparrow}, \alpha, \beta, \Gamma_{h,\omega}}{\text{Maximize}} \ \sum_{h=1}^{24}\lambda_{h}^{\rm{r}} p^{\rm{r}, \uparrow}_{h}+ \sum_{\omega=1}^{|\Omega|} \pi_{\omega}  \Bigl(\sum_{h=1}^{24}  \lambda_{h,\omega}^{\rm{b}} p^{\rm{b},\uparrow}_{h,\omega} - \notag                                                                                                                                                                                                                                                            \\  &  \hspace{3.5cm}\sum_{h=1}^{24}  \lambda_{h,\omega}^{\rm{b}} p^{\rm{b},\downarrow}_{h,\omega} - \sum_{h=1}^{24}  \lambda^{\rm{p}}s_{h,\omega} \Bigr) \label{P2:1} \\
           & \   \text{s.t.}  \  (\ref{eq:bid_constraint_relaxed}), \ \forall{h,\omega},   \label{P2:2}                                                                                                                                                                                                                                                                                                                                                                                                                                                           \\
           & \                                               \text{(\ref{P1:eq4})-(\ref{P1:eq6})}, \alpha \geq 0,   \beta \geq 0 \    \label{P2:3}                                                                                                                                                                                                                                                                                                                                                                                                                \\
           & T^{\rm{f}}_{t+1,\omega} = T^{\rm{f}}_{t,\omega} +  \frac{dt}{C^{\rm{f}}R^{\rm{cf}}} (T^{\rm{c}}_{t,\omega} - T^{\rm{f}}_{t,\omega}), \ \forall{t<J-1,\omega} \label{P2:state-space-1}                                                                                                                                                                                                                                                                                                                                                                \\
           & T^{\rm{c}}_{t+1,\omega} = T^{\rm{c}}_{t,\omega} +    \frac{dt}{C^{\rm{c}}}\Bigl(\frac{1}{R^{\rm{cf}}} (T^{\rm{f}}_{t,\omega} - T^{\rm{c}}_{t,\omega}) +  \notag                                                                                                                                                                                                                                                                                                                                                                                      \\ & \frac{1}{R^{\rm{ci}}} (T^{\rm{i}}_t - T^{\rm{c}}_{t,\omega}) - \eta  \  OD_t \ p_{h,\omega} \Bigr) + \epsilon \mathbbm{1}^{\rm{df}}_{t}, \notag \\ & \hspace{5.6cm} \forall{t<J-1,\omega} \label{P2:state-space-2} \\
           & T^{\rm{f},\text{Base}}_{t+1} = T^{\rm{f},\text{Base}}_{t} +   \frac{dt}{C^{\rm{f}}R^{\rm{cf}}} (T^{\rm{c},\text{Base}}_{t} - T^{\rm{f},\text{Base}}_{t}), \notag                                                                                                                                                                                                                                                                                                                                                                                     \\ & \hspace{6cm}\forall{t<J-1} \label{P2:state-space-3}                                                                                         \\
           & T^{\rm{c},\text{Base}}_{t+1} = T^{\rm{c},\text{Base}}_t +   \frac{dt}{C^{\rm{c}}}\Bigl(\frac{1}{R^{\rm{cf}}} (T^{\rm{f},\text{Base}}_t - T^{\rm{c},\text{Base}}_t) + \notag                                                                                                                                                                                                                                                                                                                                                                          \\ & \frac{1}{R^{\rm{ci}}} (T^{\rm{i}}_t - T^{\rm{c},\text{Base}}_t) - \eta  \  OD_t \ p_{h}^{\text{Base}} \Bigr) + \epsilon \mathbbm{1}^{\rm{df}}_{t}, \notag \\ & \hspace{6cm} \forall{t<J-1} \label{P2:state-space-4} \\
           & \ p_{h,\omega} = P^{\rm{Base}}_{h} - p^{\rm{b}, \uparrow}_{h,\omega} + p^{\rm{b}, \downarrow}_{h,\omega}, \                                                                                                  \forall{h,\omega}                                                                             \label{power:6}                                                                                                                                                                                                                           \\
        \  & p^{\rm{r}, \uparrow}_h \leq P^{\rm{Base}}_h,
        \                                                                                                                                                        \forall{h}                                                                                     \label{power:7}                                                                                                                                                                                                                                                                                   \\
        \  & p^{\rm{b}, \uparrow}_{h,\omega} \leq p^{\rm{r}, \uparrow}_h \mathbbm{1}_{h,\omega}^{\lambda^{\rm{b}}_{h,\omega} > \lambda^{\rm{s}}_{h,\omega}} , \                                                                            \forall{h,\omega}                                                                             \label{power:8}                                                                                                                                                                                                          \\
        \  & p^{\rm{b}, \uparrow}_{h,\omega} \leq u_{h,\omega}^{\uparrow} \big(P^{\rm{Base}}_{h} - P^{\rm{Min}}\big) , \                                                                                                       \forall{h,\omega}                                                                             \label{power:9}                                                                                                                                                                                                                      \\
        \  & p^{\rm{b}, \downarrow}_{h,\omega} \leq u^{\downarrow}_{h,\omega} \big(P^{\rm{Nom}} -P^{\rm{Base}}_{h}\big), \                                                                                              \forall{h,\omega}                                                                             \label{power:10}                                                                                                                                                                                                                            \\
        \  & P^{\rm{Min}} \leq p_{h,\omega} \leq P^{\rm{Nom}}, \                                                                                                                                           \forall{h,\omega}                                                                             \label{power:11}                                                                                                                                                                                                                                         \\
        \  & 0 \leq s_{h,\omega} \leq P^{\rm{Base}}_{h}, \                                                                                                                                                   \forall{h,\omega}                                                                             \label{power:12}                                                                                                                                                                                                                                       \\
        % \                                                                                             & \ p^{\rm{b}, \uparrow}_{h,\omega} + s_{h,\omega} \geq p^{\rm{r},\uparrow}_{h} \  \mathbbm{1}_{h,\omega}^{(\lambda^{\rm{bid}}_{h} < \lambda^{\rm{b}}_{h,\omega}, \lambda^{\rm{b}}_{h,\omega} > \lambda^{s}_{h})}, \                                                                                                              \forall{h,\omega} \label{power:13} \\
        \  & p^{\rm{b}, \downarrow}_{h,\omega} \geq 0.10 \  u^{\downarrow}_{h,\omega} \big(P^{\rm{Nom}} - P^{\rm{Base}}_{h}\big), \                                                                                  \forall{h,\omega}                                                                             \label{power:14}                                                                                                                                                                                                                               \\
        \  & p^{\rm{r}, \uparrow}_{h} \leq P^{\rm{Base}}_{h} \big(1 - \mathbbm{1}_{h}^{\rm{df}}\big), \                                                                                                                 \forall{h} \label{power:15}                                                                                                                                                                                                                                                                                                               \\
        \  & u_{h-1,\omega}^{\uparrow} - u_{h,\omega}^{\uparrow} + y_{h,\omega}^{\uparrow} - z_{h,\omega}^{\uparrow} = 0, \    \forall{h>1,\omega},                                                                                                         \label{aux:1}                                                                                                                                                                                                                                                                                         \\
        \  & y_{h,\omega}^{\uparrow} + z_{h,\omega}^{\uparrow} \leq 1 \                                                             \forall{h,\omega}                                                                                                                                                                     \label{aux:2}                                                                                                                                                                                                                           \\
        \  & u_{h-1,\omega}^{\downarrow} - u_{h,\omega}^{\downarrow} + y_{h,\omega}^{\downarrow} - z_{h,\omega}^{\downarrow} = 0, \                                                                                                                                                                                                                                                    \forall{h>1, \omega},                                                                                                                                        \label{aux:3} \\
        \  & y_{h,\omega}^{\downarrow} + z_{h,\omega}^{\downarrow} \leq 1 \                                                         \forall{h,\omega}                                                                                                                                                                     \label{aux:4}                                                                                                                                                                                                                           \\
        \  & u_{h,\omega}^{\uparrow} + u_{h,\omega}^{\downarrow} \leq 1 \                                                           \forall{h,\omega}                                                                                                                                                                     \label{aux:5}                                                                                                                                                                                                                           \\
        \  & y_{h,\omega}^{\uparrow} + y_{h,\omega}^{\downarrow} \leq 1 \                                                           \forall{h,\omega}                                                                                                                                                                     \label{aux:6}                                                                                                                                                                                                                           \\
        \  & z_{h,\omega}^{\uparrow} + z_{h,\omega}^{\downarrow} \leq 1 \                                                           \forall{h,\omega} \label{aux:7}                                                                                                                                                                                                                                                                                                                                                                                               \\
        \  & T^{\rm{f}}_{J,\omega} \leq T^{\rm{f}, \rm{Base}}_{J}, \ \forall{\omega} \label{temp:1}                                                                                                                                                                                                                                                                                                                                                                                                                                                               \\
        \  & y^{\downarrow}_{h, \omega} \geq z^{\uparrow}_{h, \omega}, \                                                                                                                                                                                                                                                                 \forall{h, \omega} \label{rebound:1}                                                                                                                                                                                     \\
        % \                                                                                             & \ y^{\downarrow}_{h, \omega} \leq z^{\uparrow}_{h, \omega}, \                                                                                                                                                                                                                                                                 \forall{h, \omega} \label{rebound:2}        \\
        \  & \sum_{t=4(h-1)}^{4 h} T^{\rm{f}}_{t, \omega} - T^{\rm{f}, \rm{Base}}_{t} \geq \big( z^{\downarrow}_{h, \omega} -1\big)  M,  \  \forall{h>1,\omega} \label{rebound:3}                                                                                                                                                                                                                                                                                                                                                                                 \\
        \  & \sum_{t=4(h-1)}^{4 h} T^{\rm{f}}_{t, \omega} - T^{\rm{f}, \rm{Base}}_{t} \leq \big(1 - z^{\downarrow}_{h, \omega}\big) M,  \  \forall{h>1,\omega} \label{rebound:4}                                                                                                                                                                                                                                                                                                                                                                                  \\
        \  & \sum_{k=1}^{h} y^{\downarrow}_{k,\omega} \leq y^{\uparrow}_{k, \omega}, \ \forall{h, \omega}. \label{up_reg_first}
    \end{align}
\end{subequations}

The objective function (\ref{P2:1}) maximizes the expected flexibility value of the freezer.
Constraint \eqref{P2:2} contains
(\ref{eq:bid_constraint_relaxed}), representing the McCormick relaxation of activation conditions. Recall that $\lambda^{\rm{bid}}_{h, \omega}$ in (\ref{eq:bid_constraint_relaxed}) should be replaced as defined in \eqref{eq:affine_policy}. Constraint \eqref{P2:3} declares continuous and binary variables.
%There is a set of identical constraints to (\ref{P2:2}) that simulate the baseline temperatures, $T^{\rm{f},\rm{B}}_{t}$ and $T^{\rm{c},\rm{B}}_{t}$, using the baseline power, $P^{\rm{Base}}_{h}$.. Note, the freezer specific variables are indexed by $t$, representing a time step $dt = 0.25$ whereas all other variables are indexed by hour $h$.
%from mFRR while reducing the rebound and penalty cost for all equiprobable scenarios.
%
%Equation (\ref{P2:2}) is 
%
%Equations (\ref{P2:3}-\ref{P2:5}) are the bidding policy.
%
%Constraints \eqref{power:6}-\eqref{power:15} enforce the power consumption constraints. 
%
Aligned with (\ref{eq:2ndFreezerStateSpace}), constraints \eqref{P2:state-space-1}-\eqref{P2:state-space-2} are the state-space model for the food and air temperature dynamics. Similarly, \eqref{P2:state-space-3}-\eqref{P2:state-space-4} include the baseline air temperature $T^{\rm{c},\text{Base}}_{t}$ and the   baseline food temperature $T^{\rm{f},\text{Base}}_{t}$, and model temperature dynamics for the baseline power. %These variables are used in subsequent constraints to determine when the rebound stops. 
%Note that the freezer-specific variables are indexed by $t$, representing a time step $dt = 0.25$, whereas all other variables are indexed by hour $h$. 
Recall in case the hour index $h$ runs from 1 to 24, index $t$ runs from 1 to $J=96$.
Constraint \eqref{power:6} sets the real-time power consumption $p_{h,\omega}$ equal to the baseline power $P^{\rm{Base}}_{h}$ unless there is up-regulation $p^{\rm{b}, \uparrow}_{h,\omega}$ or down-regulation $p^{\rm{b}, \downarrow}_{h,\omega}$.
Constraint (\ref{power:7}) binds the mFRR reservation $p_{h}^{\rm{r},\uparrow}$ to the baseline power.
Constraint (\ref{power:8}) ensures that up-regulation is zero when there is no need for up-regulation, and at the same time binds it to the reservation power.
Constraint (\ref{power:9}) includes the binary variable $u^{\uparrow}_{h,\omega}$, indicating whether the freezer is up-regulated in hour $h$ under scenario $\omega$. This constraint ensures that up-regulation is zero whenever $u^{\uparrow}_{h,\omega} = 0$, and otherwise restricted to the maximum up-regulation service $P^{\rm{Base}}_{h}-P^{\rm{Min}}$ that can be provided. Note that $P^{\rm{Min}}$ is the minimum consumption level of the freezer.
Constraint (\ref{power:10}) works similarly for down-regulation. Note that the binary variable $u^{\downarrow}_{h,\omega}$ indicates whether down-regulation happens, whereas $P^{\rm{Nom}}$ is the nominal (maximum) consumption level of the freezer.
Constraint (\ref{power:11}) restricts the power consumption to lie within the minimum and nominal rates.
Constraint (\ref{power:12}) binds the slack variable $s_{h,\omega}$, representing the service not delivered as promised.
Constraint (\ref{power:14}) ensures that down-regulation is equal to at least 10\% of the down-regulation capacity.
Constraint (\ref{power:15}) prohibits any up-regulation when defrosting occurs.
Constraints \eqref{aux:1}-\eqref{aux:7} define auxiliary binary variables $y^{\uparrow}_{h,\omega}$, $y^{\downarrow}_{h,\omega}$, $z^{\uparrow}_{h,\omega}$, and $z^{\downarrow}_{h,\omega}$, identifying transitions from/to up-regulation and down-regulation.
During all hours with up-regulation, $y^{\uparrow}_{h,\omega}=1$. In the hour that up-regulation is stopped, $z^{\uparrow}_{h,\omega}$ is 1. There is a similar definition for $y^{\downarrow}_{h,\omega}$ and $z^{\downarrow}_{h,\omega}$ related to down-regulation.
See Chapter 5 of  \cite{morales2013integrating} for complete details.
Constraint \eqref{temp:1} restricts the food temperature for the last time period $J$.
%For every time period $t$, \eqref{temp:2} and \eqref{temp:3} bind the air temperature by enforcing the maximum deviation $\Delta^{\rm{max}}$.
%
Constraints \eqref{rebound:1}-\eqref{rebound:4} control the rebound behavior such that the rebound finishes when the temperature is below the baseline temperature. Note that $M$ is a sufficiently big positive constant such that the food temperature is allowed to deviate from the baseline. Also, they ensure that the rebound happens right after up-regulation.
Lastly, (\ref{up_reg_first}) ensures that up-regulation happens first. This makes sense since it impossible (or at least difficult) to anticipate potential up-regulation events in the power system. As such, it does not make sense to pre-cool (or pre-heat) a TCL in the context of mFRR.

\section*{\textcolor{black}{Appendix B: Sensitivity analysis of scenarios}}\label{appendix:B}

\textcolor{black}{A sensitivity analysis was carried out to investigate the number of scenarios to use in the mFRR model using  2021 data. Fig. \ref{fig:admm_sensitivity} shows the in-sample (IS) and out-of-sample (OOS) costs when increasing the number of scenarios used from 2021. It was eventually chosen to use 50 scenarios as this provided a satisfactory performance both IS and OOS --- a higher number of scenarios did not yield any particular improvement.}

\begin{figure}[!t]
    \centering
    \includegraphics[width=\columnwidth]{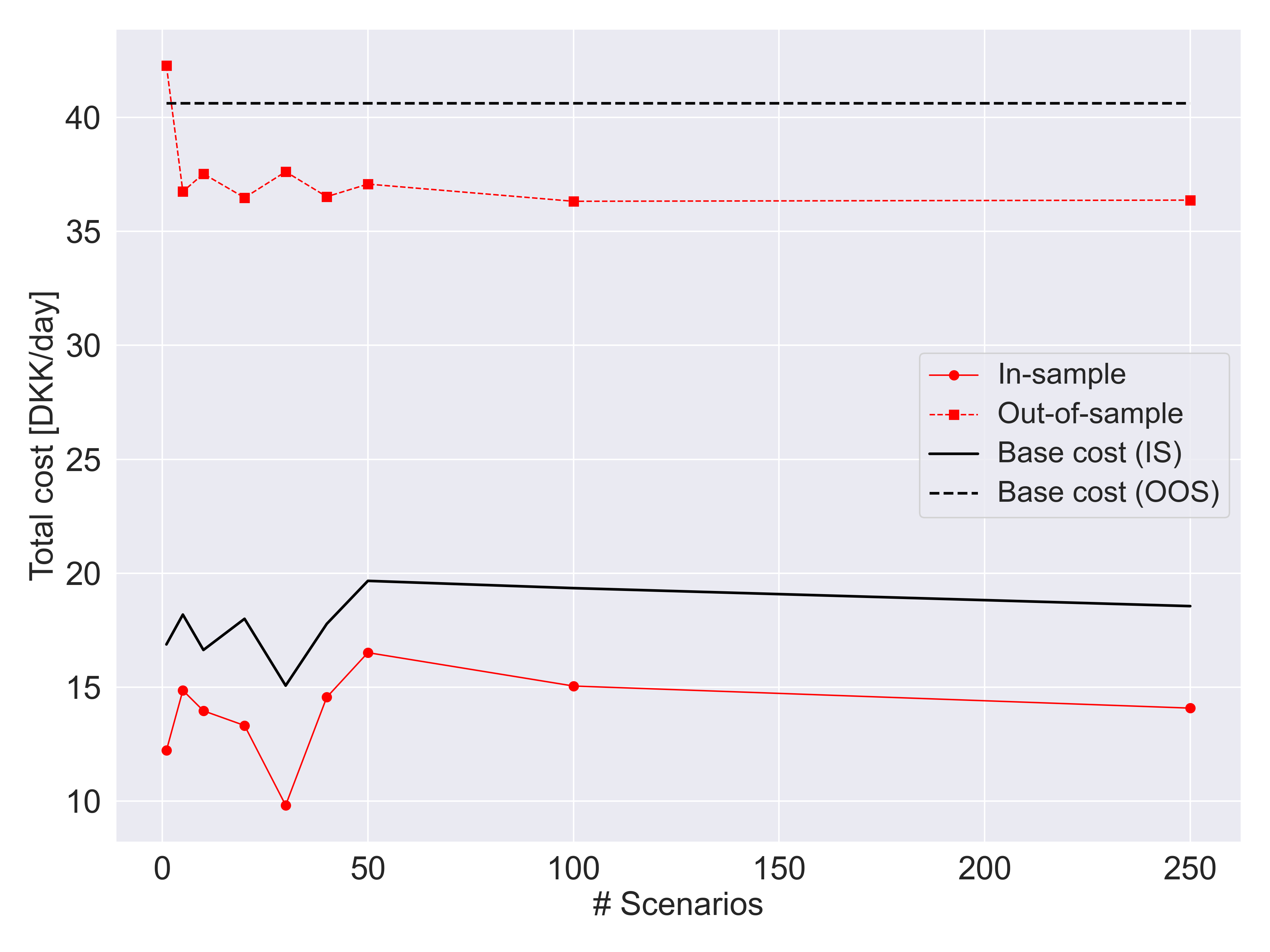}
    \caption{\textcolor{black}{Total cost per day for supermarket freezer using the mFRR model in \eqref{P1:compact_model} with 2021 data and the ADMM solution strategy described in Section \ref{sec:admm} and for a different number of scenarios used from 2021.}}
    \label{fig:admm_sensitivity}
\end{figure}

% \pagebreak
%\bibliographystyle{model5-names}\biboptions{authoryear}
%\bibliography{bibliography/Bibliography}
\printbibliography

\end{document}